\documentclass[pdflatex,sn-mathphys]{sn-jnl}

\usepackage{siunitx}
\usepackage{array,booktabs}
\usepackage{csvsimple}
\usepackage{xspace}
\usepackage[graphicx]{realboxes}
\PassOptionsToPackage{hyphens,sloppy}{url}\usepackage{hyperref}

\newcommand{\printsparsedata}[1]{{
\begin{center}
\small
\csvreader[tabular={l|S[table-format=1.2e3]|S[table-format=1.2e3]|S[table-format=1.2e3]|S[table-format=1.2e3]|S[table-format=1.2e3]|S[table-format=1.2e3]|@{}c}, table head= \bfseries SRQ & \ensuremath{P_{10}(\bar x)} & \ensuremath{P_{50}(\bar x)} & \ensuremath{P_{90}(\bar x)} & \ensuremath{P_{10}(\sigma)} & \ensuremath{P_{50}(\sigma)} & \ensuremath{P_{90}(\sigma)} & \\\midrule, table foot=\bottomrule]{sparse_data/#1.tex}%
{}%
{\csvcoli & \csvcolii & \csvcoliii & \csvcoliv & \csvcolv & \csvcolvi & \csvcolvii &}
\normalsize
\end{center}
}}

\newcommand{\printdistances}[1]{{
\begin{center}
\footnotesize
\scalebox{0.9}{
\csvreader[tabular={l|S[table-format=1.2e3]|S[table-format=1.2e3]|S[table-format=1.2e3]|S[table-format=1.2e3]|S[table-format=1.2e3]|S[table-format=1.2e3]|S[table-format=1.2e3]|S[table-format=1.2e3]|@{}c}, table head=\textbf{dist [m]} & \text{Austin} & \text{CSIRO} & \text{DARSim} & \text{DARTS} & \text{LANL} & \text{Melbourne} & \text{Stanford} & \text{Stuttgart} & \\\midrule, table foot=\bottomrule]{distances/#1.tex}%
{}%
{\csvcoli & \csvcolii & \csvcoliii & \csvcoliv & \csvcolv & \csvcolvi & \csvcolvii & \csvcolviii & \csvcolix &}
}
\normalsize
\end{center}
}}

\definecolor{mplc0}{HTML}{1f77b4}
\definecolor{mplc1}{HTML}{ff7f0e}
\definecolor{mplc2}{HTML}{2ca02c}
\definecolor{mplc3}{HTML}{d62728}
\definecolor{mplc4}{HTML}{9467bd}
\definecolor{mplc5}{HTML}{8c564b}
\definecolor{mplc6}{HTML}{e377c2}
\definecolor{mplc7}{HTML}{7f7f7f}
\definecolor{mplc8}{HTML}{bcbd22}
\definecolor{mplc9}{HTML}{17becf}

\newcommand{\cotwo}{CO\textsubscript{2}\xspace}
\newcommand{\austin}{{\color{mplc0}\texttt{Austin}}\xspace}
\newcommand{\csiro}{{\color{mplc1}\texttt{CSIRO}}\xspace}
\newcommand{\darsim}{{\color{mplc2}\texttt{Delft-DARSim}}\xspace}
\newcommand{\darts}{{\color{mplc3}\texttt{Delft-DARTS}}\xspace}
\newcommand{\hw}{{\color{mplc4}\texttt{Heriot-Watt}}\xspace}
\newcommand{\lanl}{{\color{mplc6}\texttt{LANL}}\xspace}
\newcommand{\melbourne}{{\color{mplc7}\texttt{Melbourne}}\xspace}
\newcommand{\stanford}{{\color{mplc8}\texttt{Stanford}}\xspace}
\newcommand{\stuttgart}{{\color{mplc9}\texttt{Stuttgart}}\xspace}
\newcommand{\coordus}{\texttt{CoordUS}\xspace}
\newcommand{\expub}{\texttt{ExpUB}\xspace}

\jyear{2022}%

\begin{document}

\title[The FluidFlower International Benchmark Study]{The FluidFlower International Benchmark Study: Process, Modeling Results, and Comparison to Experimental Data}

\author*[1]{\fnm{Bernd} \sur{Flemisch}}\email{bernd@iws.uni-stuttgart.de}

\author[2,3]{\fnm{Jan M.} \sur{Nordbotten}}\email{jan.nordbotten@uib.no}

\author[4]{\fnm{Martin} \sur{Fernø}}\email{martin.ferno@uib.no}

\author[5]{\fnm{Ruben} \sur{Juanes}}\email{juanes@mit.edu}

\author[1]{\fnm{Holger} \sur{Class}}\email{holger.class@iws.uni-stuttgart.de}

\author[6]{\fnm{Mojdeh} \sur{Delshad}}\email{delshad@mail.utexas.edu}

\author[7]{\fnm{Florian} \sur{Doster}}\email{f.doster@hw.ac.uk}

\author[8]{\fnm{Jonathan} \sur{Ennis-King}}\email{jonathan.ennis-king@csiro.au}

\author[9]{\fnm{Jacques} \sur{Franc}}\email{jfranc@stanford.edu}

\author[7,10]{\fnm{Sebastian} \sur{Geiger}}\email{s.geiger@tudelft.nl}

\author[1]{\fnm{Dennis} \sur{Gläser}}\email{dennis.glaeser@iws.uni-stuttgart.de}

\author[8]{\fnm{Christopher} \sur{Green}}\email{chris.green@csiro.au}

\author[8]{\fnm{James} \sur{Gunning}}\email{james.gunning@csiro.au}

\author[10]{\fnm{Hadi} \sur{Hajibeygi}}\email{h.hajibeygi@tudelft.nl}

\author[8]{\fnm{Samuel J.} \sur{Jackson}}\email{samuel.jackson@csiro.au}

\author[6]{\fnm{Mohamad} \sur{Jammoul}}\email{jammoul@utexas.edu}

\author[12]{\fnm{Satish} \sur{Karra}}\email{satkarra@lanl.gov}

\author[9]{\fnm{Jiawei} \sur{Li}}\email{jiaweili@stanford.edu}

\author[13]{\fnm{Stephan K.} \sur{Matthäi}}\email{stephan.matthai@unimelb.edu.au}

\author[12]{\fnm{Terry} \sur{Miller}}\email{tamiller@lanl.gov}

\author[13]{\fnm{Qi} \sur{Shao}}\email{shao.q@unimelb.edu.au}

\author[9]{\fnm{Catherine} \sur{Spurin}}\email{cspurin@stanford.edu}

\author[12]{\fnm{Philip} \sur{Stauffer}}\email{stauffer@lanl.gov}

\author[9]{\fnm{Hamdi} \sur{Tchelepi}}\email{tchelepi@stanford.edu}

\author[10]{\fnm{Xiaoming} \sur{Tian}}\email{x.tian-1@tudelft.nl}

\author[12]{\fnm{Hari} \sur{Viswanathan}}\email{viswana@lanl.gov}

\author[10]{\fnm{Denis} \sur{Voskov}}\email{d.v.voskov@tudelft.nl}

\author[10]{\fnm{Yuhang} \sur{Wang}}\email{y.wang-25@tudelft.nl}

\author[10]{\fnm{Michiel} \sur{Wapperom}}\email{m.b.wapperom@tudelft.nl}

\author[6]{\fnm{Mary F.} \sur{Wheeler}}\email{mfw@ices.utexas.edu}

\author[]{\fnm{Andrew} \sur{Wilkins$^\text{14}$}}\email{andrew.wilkins@csiro.au}

\author[13]{\fnm{AbdAllah A.} \sur{Youssef}}\email{abdallahy@student.unimelb.edu.au}

\author[10]{\fnm{Ziliang} \sur{Zhang}}\email{z.zhang-15@tudelft.nl}


\affil*[1]{\orgdiv{Department of Hydromechanics and Modelling of Hydrosystems}, \orgname{University of Stuttgart}, \orgaddress{\city{Stuttgart}, \country{Germany}}}

\affil[2]{\orgdiv{Center for Modeling of Coupled Subsurface Dynamics, Department of Mathematics}, \orgname{University of Bergen}, \orgaddress{\city{Bergen}, \country{Norway}}}

\affil[3]{\orgdiv{Center of Sustainable Subsurface Resources}, \orgname{Norwegian Research Center}, \orgaddress{\city{Bergen}, \country{Norway}}}

\affil[4]{\orgdiv{Department of Physics and Technology}, \orgname{University of Bergen}, \orgaddress{\city{Bergen}, \country{Norway}}}

\affil[5]{\orgdiv{Department of Civil and Environmental Engineering}, \orgname{Massachusetts Institute of Technology}, \orgaddress{\city{Cambridge}, \state{Massachusetts}, \country{USA}}}

\affil[6]{\orgdiv{Center for Subsurface Modeling, Oden Institute for Computational Engineering and Sciences}, \orgname{The University of Texas at Austin}, \orgaddress{\city{Austin}, \state{Texas}, \country{USA}}}

\affil[7]{\orgdiv{Institute of Geoenergy Engineering}, \orgname{Heriot Watt University}, \orgaddress{\city{Edinburgh}, \country{UK}}}

\affil[8]{\orgname{CSIRO Energy}, \orgaddress{\city{Clayton North}, \country{Australia}}}

\affil[9]{\orgdiv{Department of Energy Science and Engineering}, \orgname{Stanford University}, \orgaddress{\city{Stanford}, \state{California}, \country{USA}}}

\affil[10]{\orgdiv{Department of Geoscience and Engineering}, \orgname{Delft University of Technology}, \orgaddress{\city{Delft}, \country{The Netherlands}}}

\affil[12]{\orgdiv{Computational Earth Science}, \orgname{Los Alamos National Laboratory}, \orgaddress{\city{Los Alamos}, \state{New Mexico}, \country{USA}}}

\affil[13]{\orgdiv{Peter Cook Center for CCS Research \& Department of Infrastructure Engineering}, \orgname{The University of Melbourne}, \orgaddress{\city{Parkville}, \country{Australia}}}

\affil[14]{\orgdiv{CSIRO Mineral Resources}, \orgname{Queensland Centre for Advanced Technologies}, \orgaddress{\city{Kenmore}, \country{Australia}}}


\abstract{Successful deployment of geological carbon storage (GCS) requires an extensive use of reservoir simulators for screening, ranking and optimization of storage sites. However, the time scales of GCS are such that no sufficient long-term data is available yet to validate the simulators against. As a consequence, there is currently no solid basis for assessing the quality with which the dynamics of large-scale GCS operations can be forecasted.

To meet this knowledge gap, we have conducted a major GCS validation benchmark study. To achieve reasonable time scales, a laboratory-size geological storage formation was constructed (the ``FluidFlower''), forming the basis for both the experimental and computational work. A validation experiment consisting of repeated GCS operations was conducted in the FluidFlower, providing what we define as the true physical dynamics for this system.  Nine different research groups from around the world provided forecasts, both individually and collaboratively, based on a detailed physical and petrophysical characterization of the FluidFlower sands.

The major contribution of this paper is a report and discussion of the results of the validation benchmark study, complemented by a description of the benchmarking process and the participating computational models. The forecasts from the participating groups are compared to each other and to the experimental data by means of various indicative qualitative and quantitative measures. By this, we provide a detailed assessment of the capabilities of reservoir simulators and their users to capture both the injection and post-injection dynamics of the GCS operations.}


\maketitle

\newpage
\section{Introduction}
\label{sec:introduction}

Geological carbon storage (GCS) has the potential to close the gap between CO$_2$ emissions from legacy carbon-based power sources and the required emission reductions as outlined in the IPCC reports~\cite{Bachu:2007:CSC,Pacala:2004:SWS,Halland:2013:CSA,Metz:2005:ISR}. Furthermore, GCS can play a role in negative emissions strategies in combination with biofuels~\cite{Johnson:2014:NBC}, and in the production of so-called ``blue hydrogen''~\cite{Noussan:2021:RGB}. In order to realize this potential in a safe and cost-efficent manner, large scale deployment of GCS relies heavily on modeling and numerical simulation studies to assess the suitability of potential geological formations (predominantly subsurface aquifers). Such modeling studies have been heavily relied upon in existing assessments of storage potential~\cite{Juanes:2010:FCP,Lindeberg:2009:DCS,Kopp:2009:ICSDA,Kopp:2009:ICSES,Niemi:2016:HES,Sharma:2011:COP}. The generation of simulation-based data and knowledge in fields like GCS with huge societal impact eventually requires communication to political decision makers. Transparent simulation work flows, reproducibility of data and increased confidence in simulation results, e.g. as a result of comprehensive benchmarking, are key factors for communication or a participation of stakeholders in the modeling process \cite{Scheer:2021:SEM}.

On the other hand, only a few dozen large-scale carbon storage operations are currently active globally~\cite{Steyn:2022:GSC}, and of these, none are in a post-injection phase following a multi-decadal injection period. As such, the modeling and simulation community does not have a robust data set to assess their forecasting skill, and significant uncertainty is associated with our ability to accurately capture the dominant physical processes associated with GCS. Pilot studies provide some measure of information~\cite{Preston:2005:IGW,Lueth:2020:GMI}, yet the fundamental nature of the subsurface means that the data collected will always be relatively sparse, in particular spatially. As a partial remedy to this, several code comparison studies have been conducted~\cite{Pruess:2004:CIB,Class:2009:BSP,Nordbotten:2012:UPS}. However, none of these studies were conducted in the presence of a physical ground truth.

This study aims to provide a first assessment of the predictive skills of the GCS modeling and simulation community. To achieve this goal, we are exploiting the newly constructed ``FluidFlower'' experimental facility at the University of Bergen. Within this experimental rig, a geological model with characteristic features from the Norwegian Continental Shelf was constructed. Initial geological and petrophysical characterization was completed, together with a single-phase tracer test. With this basis, we conducted a double-blind study: On one hand, laboratory scale GCS was repeatedly conducted and measured at the University of Bergen, where the corresponding group will be labelled as \expub in the following. On the other hand, academic research groups active in GCS around the world were invited to participate in a forecasting study, coordinated by the University of Stuttgart, in the following indicated by \coordus. Aided by the fact that the pandemic reduced academic travel, we were able to fully ensure that no physical interaction was present between the participating groups, and all digital communication was restricted, moderated, and archived to ensure the integrity of the double-blind study. As detailed in the following, the participants of the forecasting study were both asked to provide independent forecasts, and then subsequently invited to update their forecasts in view of group interactions. 

In this contribution, we report the final results of the comparison study, emphasizing 1) The degree of correlation between forecasts from the diverse set of participating groups, and 2) The degree of correlation between the forecasts and the measurements from the laboratory scale GCS conducted in the FluidFlower. Seen together, this provides both a measure of repeatability among forecasts (seen from an operational perspective), and also an indication of forecasting skill.   

We structure the paper as follows. Section \ref{sec:description} introduces some basic required terminology, describes the validation experiment, and illustrates the benchmarking process. The participating groups and corresponding models are introduced in Section \ref{sec:groups}. In Section \ref{sec:modeling_results}, the modeling results are presented and discussed by means of qualitative and quantitative assessments. Section \ref{sec:comparison} provides a comparison of the modeling results with the experimental data. A conclusion and an outlook are given in Section \ref{sec:conclusion}.


\section{Benchmarking Methodology}
\label{sec:description}

We start this section by introducing some fundamental concepts and terminology based on \cite{Oberkampf:2010:VVS,ASME:2006:GVV}. While the term \textit{verification} describes ``the process of determining that a computational model accurately represents the underlying mathematical model and its solution'', \textit{validation} refers to ``the process of determining the degree to which a model is an accurate representation of the real world from the perspective of the intended uses of the model''. In addition, \emph{calibration} is the process of adjusting parameters in the computational model to improve agreement with data.

A \emph{validation experiment} like the one presented below in Section \ref{sec:experiment} is ``designed, executed, and analyzed for the purpose of quantitatively determining the ability of a mathematical model expressed in computer software to simulate a well-characterized physical process''.
As described in further detail below in Section \ref{sec:process}, we perform a \emph{validation benchmark}~\cite{Oberkampf:2008:VVB,Oberkampf:2010:VVS}, where the experiment provides measured data against which the simulation results are to be compared.

\subsection{The validation experiment}
\label{sec:experiment}
\begin{figure}[hbt]
\centering
\includegraphics[width=0.95\textwidth]{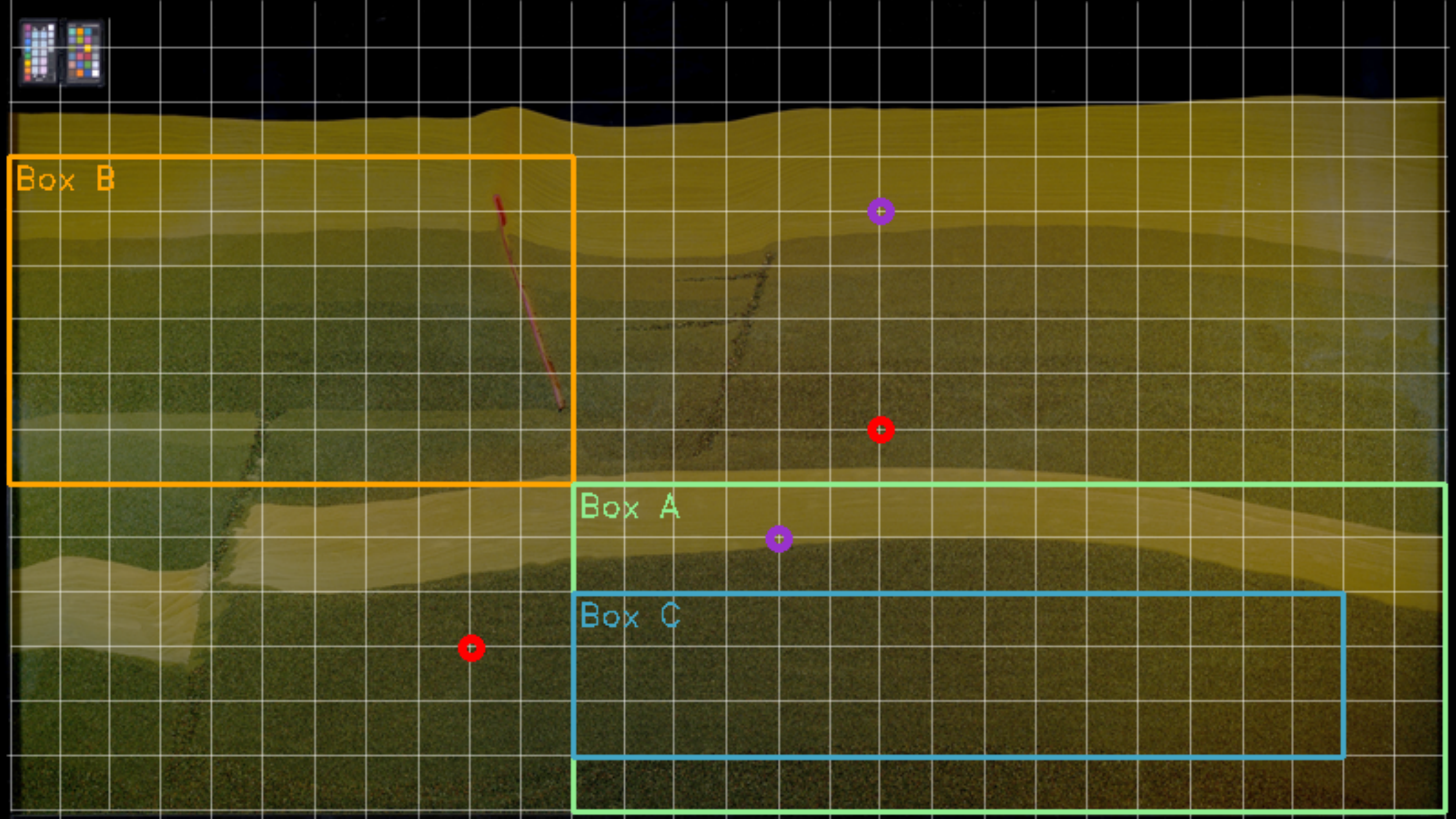}
\caption{Photograph image of the benchmark geometry with overlaid laser grid \cite[Figure 8]{Nordbotten:2022:FBD}. The brightest facies are the fine-sand barriers. The red circles indicate the injection points, while the purple circles depict the pressure sensors. Boxes A-C correspond to regions for the evaluation of different system response quantities.}
\label{fig:benchmark_geometry}
\end{figure}

In the following, we provide a very brief description of the experiment performed with the FluidFlower rig. For details, we refer to the original benchmark description \cite{Nordbotten:2022:FBD} and the experimental paper \cite{Ferno:2023:MCI}.
Figure \ref{fig:benchmark_geometry} shows the geometrical setup where the rig has been filled with sand of six different types to build up several layers of varying permeability, including three fault-like structures. 
Initially, the pore space was fully water-saturated and the top of the water table was subject to atmospheric conditions in terms of pressure and temperature\footnote{Obviously, the density difference between \cotwo and water is much greater at this low pressure than it is at typical reservoir depths. We will address the relevance of our study for realistic GCS scenarios in Section \ref{sec:conclusion}.}. Gaseous \cotwo was injected over a total of five hours by means of two injection ports. The distribution of \cotwo throughout the rig was monitored over five days after the injection start. In total, five experimental runs were performed between November 2021 and January 2022. The experimental team \expub tried to establish identical operational conditions during the runs.

The description of the experimental setup in \cite{Nordbotten:2022:FBD} addressed the external geometry, stratification, facies properties, faults, fluid properties, operational conditions and well test data. In particular, the stratification was described by high-resolution photographs, from which the participating groups had to determine the location of the different sand layers. This was complemented by details on the sedimentation process and pre-injection flushing procedures. Concerning the facies, information was provided on grain size distributions as well as on measurements of absolute permeability, porosity, relative permeability endpoints and capillary entry pressures. The purpose of the well test data was to allow for calibration of the numerical models. In particular, the provided pressure\footnote{The injection pressures were reported at a sensor that was separated from the injection point by the length of small diameter tubing. Taking the pressure drop along that tubing into account influences the result of the calibration.} and tracer flow data could be employed to estimate the permeability distribution over the different facies.

The description also defined the \emph{System Response Quantities} (SRQs) which should be reported by the benchmark participants. The individual SRQs will be introduced in detail in Section \ref{sec:modeling_results}.

\subsection{Benchmarking Process}
\label{sec:process}

Table \ref{tab:chronology} shows the chronology of the benchmark process.
After a common preparation phase for finalizing the benchmark description \cite{Nordbotten:2022:FBD}, a so-called blind phase of three months started, where there was no direct communication between different participating groups or with \expub allowed. 
\begin{table}[hbt]
\begin{center}
\begin{tabular}{|p{0.2\textwidth}|p{0.69\textwidth}|}\hline
30.04.2021 & Closed call for participation opens. \\\hline
15.06.2021 & Call closes.\\\hline
15.07.2021 & Preliminary benchmark description supplied to participants.\\\hline
16.07.--19.08.2021 & Preparation phase, discussion possible among all participants and \expub.\\\hline
20.08.2021 & Deadline for feedback on preliminary benchmark description.\\\hline
16.09.2021 & Kick-off Zoom meeting, second iteration of benchmark description distributed.\\\hline
17.09.--08.10.2021 & Open discussion for finalizing the description.\\\hline
08.10.2021 & Final benchmark description circulated to participants.\\\hline
09.10.2021--11.01.2022 & Blind phase, no direct communication between different participants or with \expub.\\\hline
09.01.2022 & Deadline for submitting blind benchmark data.\\\hline
12.01.2022 & Virtual workshop and comparison of ``fully blind'' simulation forecasts.\\\hline
12.01.--25.04.2022 & Synchronization phase, communication between all participants enabled, but not with \expub.\\\hline
22.04.2022 & Deadline for submitting final benchmark data.\\\hline
26.--28.04.2022 & Workshop in Norway with presentation of final simulation forecasts, experimental results, model calibration study, and synthesis of results.\\\hline
\end{tabular}\\[2mm]
\end{center}
\caption{Chronology of the FluidFlower benchmark process.}
\label{tab:chronology}
\end{table}
All upcoming issues of the modelers were directed to \coordus and potentially anonymously forwarded to \expub. After agreeing on an answer between \coordus and \expub, that answer was either broadcasted to all participating groups or given to the questioner only. At the end of the blind phase, each participating group provided initial forecasts to \coordus. This was followed by a first meeting of all participating groups where the results were revealed and discussed, still without any involvement of \expub. This meeting initiated a so-called synchronization phase of another three months, allowing the forecasting groups to learn from each other's work and bring this knowledge into their own forecasts. In particular, the synchronization phase included two more common participant meetings. At its end, final forecasts were recorded before an in-person workshop outside of Bergen, Norway, where forecasts and experiments were compared for the first time.

In order to protect the integrity of the results, dedicated communication rules were followed during the
different phases of the benchmarking process. To facilitate remote communication between
participants, and also to store this communication for evaluating the benchmarking process, a Discord
server was set up\footnote{\url{https://discord.gg/8Q5fZS3T47}}. Apart from a general channel that was initially open to everyone involved, a private channel was installed for each participating group which could be used for communicating with the benchmark organizers.

All result data was uploaded by the participants to Git repositories within a GitHub organization ``FluidFlower''\footnote{\url{https://github.com/fluidflower}}. Each participating group got write access to a dedicated repository named after their institution. During the blind phase, only the participants themselves had access to their respective repositories. For the synchronization phase, read access to all participant repositories was granted for all participants. After the workshop in April, the repositories were opened further to include also the results from the physical experiments. Upon submission of this paper, the relevant repositories have been turned public.

\section{Participating Groups and Models}
\label{sec:groups}

In total, nine groups, each consisting of two to five individuals, participated in the FluidFlower benchmark study. In the following, they are indicated by the location or name of the corresponding institution as \austin, \csiro, \darsim, \darts, \hw, \lanl, \melbourne, \stanford and \stuttgart.
Table \ref{tab:choices} lists relevant modeling choices of the participating groups.
\begin{table}[hbt]
\centering
\scalebox{0.9}{
\Rotatebox{-90}{%
\begin{tabular}{|p{0.09\textheight}|p{0.09\textheight}|p{0.09\textheight}|p{0.09\textheight}|p{0.09\textheight}|p{0.09\textheight}|p{0.09\textheight}|p{0.09\textheight}|p{0.09\textheight}|p{0.09\textheight}|}
\textbf{Type} & \austin & \csiro & \darsim & \darts & \hw & \lanl & \melbourne & \stanford & \stuttgart\\\midrule
\textbf{PDEs} & CMB & CMB & CMB & CMB & pseudo black oil & CMB & pseudo black oil + transport of dissolved \cotwo & pseudo black oil & CMB\\\midrule
$p_\text{c}$, $k_\text{r}$ & BC & BC & BC & const. $p_\text{E}$, power law & powerlaw for rel perms, BC for fine sands, vG for coarse sands & linear & BC & BC & BC \\\midrule
\textbf{EOS} & \cite{Peng:1976:NTE} & \cite{spycher2003,spycher2005} & \cite{spycher2003} & liquid: \cite{Ziabakhsh:2012:EST}, gas: \cite{Peng:1976:NTE} &   & \cite{Duan:2003:IMC} & \cite{spanwagner1996,Span:2003:EST} & \cite{Weiss:1974:CDW,Sandve:2021:CDF,Duan:2003:IMC} & \cite{spycher2005,Duan:2003:IMC,spycher2003} \\\midrule
\textbf{Density} & \cite{Peng:1976:NTE} & liquid: \cite{iapws1997,garcia2001}, gas: \cite{spanwagner1996}
& $\varrho_l = (\varrho_b^\text{STC} + \varrho_\text{CO2}^\text{STC}R_S)/B_b$, \cite{Soave:1972:ECM}
& liquid: \cite{iapws1997,garcia2001}, gas: \cite{Peng:1976:NTE}
& liquid: exp. with $p$, linear with \cotwo conc., gas: exp. with $p$
& \cite{spanwagner1996}
& Derived from miscibility data reported in \cite{Carroll:1991:SCD} & \cite{Fenghour:1998:VCD,spanwagner1996}
& \cite{iapws1997,spanwagner1996} \\\midrule
\textbf{Solubility limit [\unit{\kilogram\per\cubic\meter}]} & 1.496 & 1.786 & 1.649 & 1.9 & 2.0 & 2.0 & $f(p)$, \numrange{1.752}{2.0093} & 1.5 & 1.845 \\\midrule
\textbf{Domain volume [\unit{\cubic\meter}]} & 9.1e-2 & 8.65e-2 & 8.3e-2 & 9.2e-2 & 8.4e-2 & 8.4e-2 & 8.18e-2 & 8.4e-2 & 8.75e-2 \\\midrule
\textbf{Disc.} & MFEM & CC-FV & CC-FV & CC-FV & CC-FV & CC-FV & DFEFVM & CC-FV & CC-FV \\\midrule
\textbf{\# cells} & 9,100 & 44,284 & 43,758 & 48,274 & 42,000 & 42,000 & 14,822 & 6,094 / 21,392 & 26,099 \\\midrule
\textbf{Software} & IPARS & MOOSE~\cite{wilkins21} & DARSim \cite{Wang:2022:AHT} & DARTS \cite{Lyu:2021:OLA} & MRST-2021b \cite{Lie:2019:IRS} & FEHM \cite{Zyvoloski:1997:SMM}, PFLOTRAN \cite{Lichtner:2015:PUM} & CSMP++ \cite{Matthai:2001:CSP} & AD-GPRS \cite{Zhou:2013:SML,Garipov:2018:UTC,Zhou:2012:PGP,Younis:2010:ALC} & DuMu$^\text{x}$ \cite{Koch:2021:DOS} \\\midrule
\end{tabular}
}
}
\caption{Modeling choices of the participating groups. Used abbreviations: component mass balances ``CMB'', phase mass balances ``PMB'', Brooks--Corey ``BC'', mixed finite elements ``MFEM'', cell-centered finite volumes ``CC-FV'', collocated finite-element finite-volume method with embedded discontinuities ``DFEFVM''.}
\label{tab:choices}
\end{table}

In terms of the partial differential equations constituting the main part of the mathematical model, almost all participants employ component mass balances. Apart from two exceptions \austin and \melbourne, the choice of spatial discretization is uniform with cell-centered finite volumes. All groups but \melbourne employ a standard implicit Euler time discretization and solve the resulting discrete equations in a fully-coupled fully-implicit manner. Things start to differ more when it comes to the constitutive relations. While the majority of the participants uses Brooks--Corey relationships for the capillary pressure and relative permeability, also other approaches such as linear relationships are employed. Moreover, various equations of state for determining the phase compositions as well as the phase densities are considered. Additionally to these principal choices, the participating computational models differ in their employed spatial parameters such as the assumed intrinsic permeabilities, porosities, residual saturations and others. These parameters may depend on the considered sand type, i.e., on the spatial location. They have been collected for each participating group in a file \texttt{spatial\_parameters.csv} in the top level of the respective GitHub repository.

\clearpage

\section{Modeling Results}
\label{sec:modeling_results}

In the following, we provide and discuss the modeling results which are requested in form of SRQs by the benchmark description. They are grouped into three categories: dense data spatial maps in Section \ref{sec:results_spatial_maps}, dense data time series in Section \ref{sec:results_time_series}, and sparse data in Section \ref{sec:results_sparse_data}.

\subsection{Dense data spatial maps}
\label{sec:results_spatial_maps}

The participants were asked to provide snapshots of the spatial phase distribution at 24, 48, 72, 96 and \qty{120}{\hour} after injection start, particularly, the saturation of gaseous \cotwo as well as the concentration of \cotwo in the liquid phase. While each participating group was free to define the computational grid for performing simulations, results should be reported on a uniform grid consisting of \qty{1}{\cm} by \qty{1}{\cm} cells.

\subsubsection{Saturation}

\begin{figure}[hbt]
\centering
\includegraphics[width=0.99\textwidth]{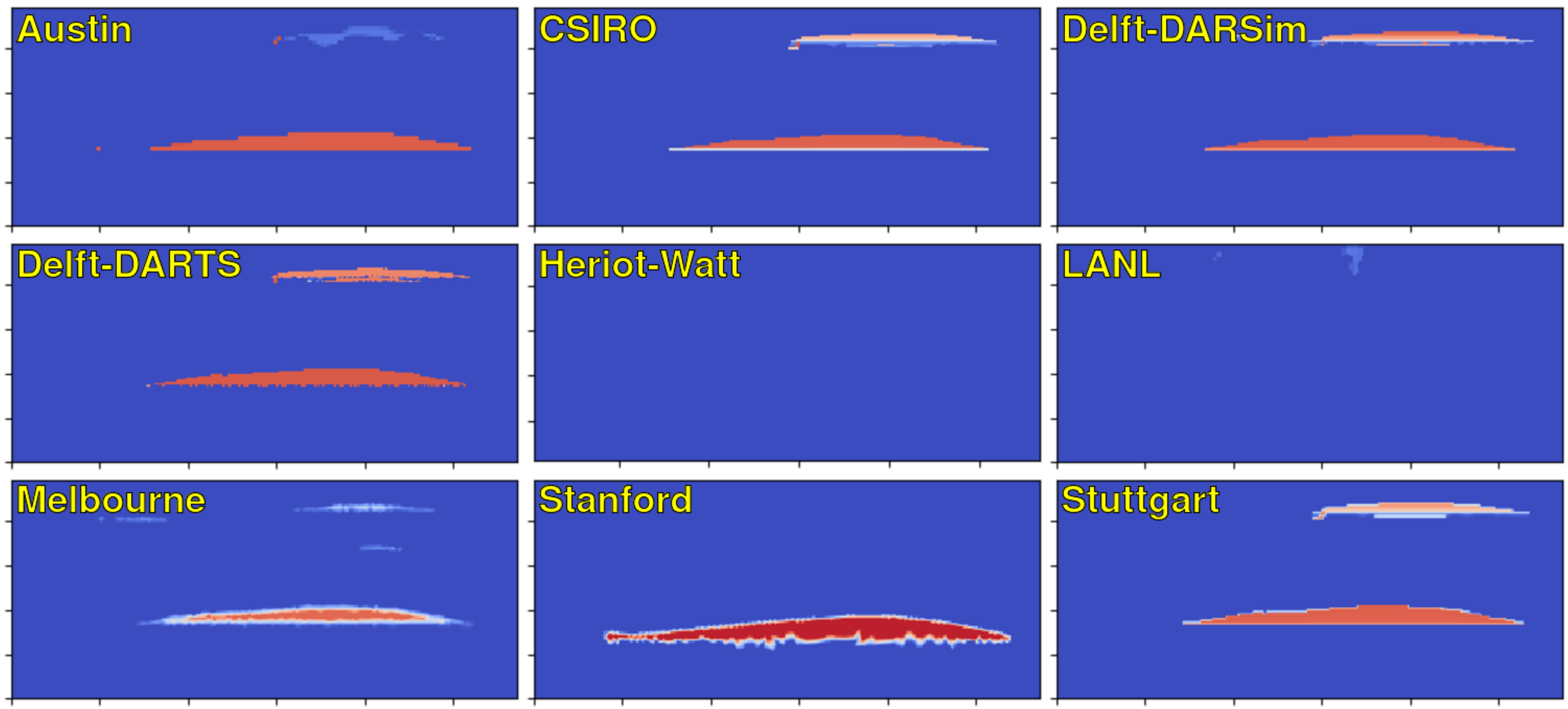}
\caption{Spatial distribution of gaseous \cotwo after \qty{24}{\hour}. The minimum for the color map is at 0 \cotwo saturation indicated by blue, the maximum at 1 indicated by red.}
\label{fig:saturation_one}
\end{figure}

\begin{figure}[hbt]
\centering
\includegraphics[width=0.99\textwidth]{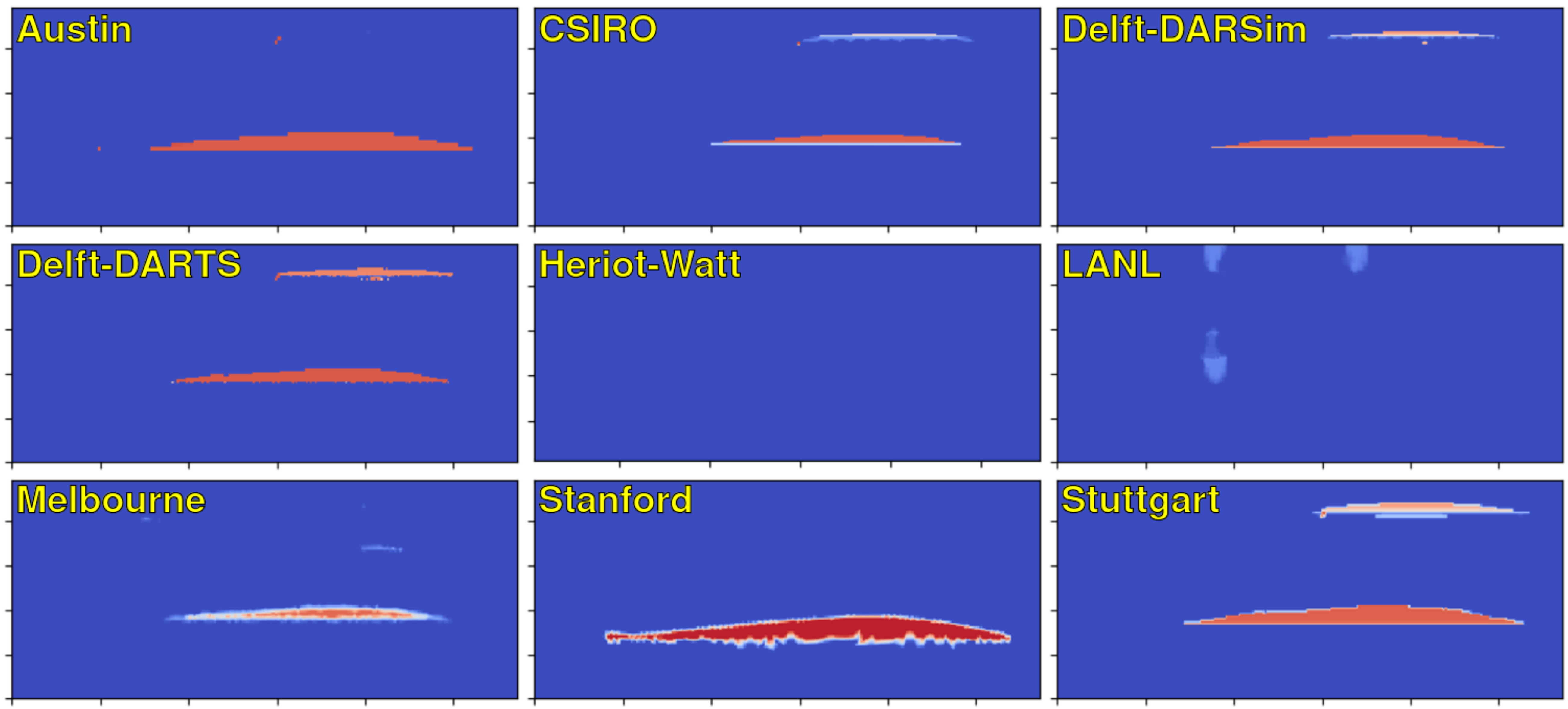}
\caption{Spatial distribution of gaseous \cotwo after \qty{48}{\hour}. The minimum for the color map is at 0 \cotwo saturation indicated by blue, the maximum at 1 indicated by red.}
\label{fig:saturation_two}
\end{figure}

\begin{figure}[hbt]
\centering
\includegraphics[width=0.99\textwidth]{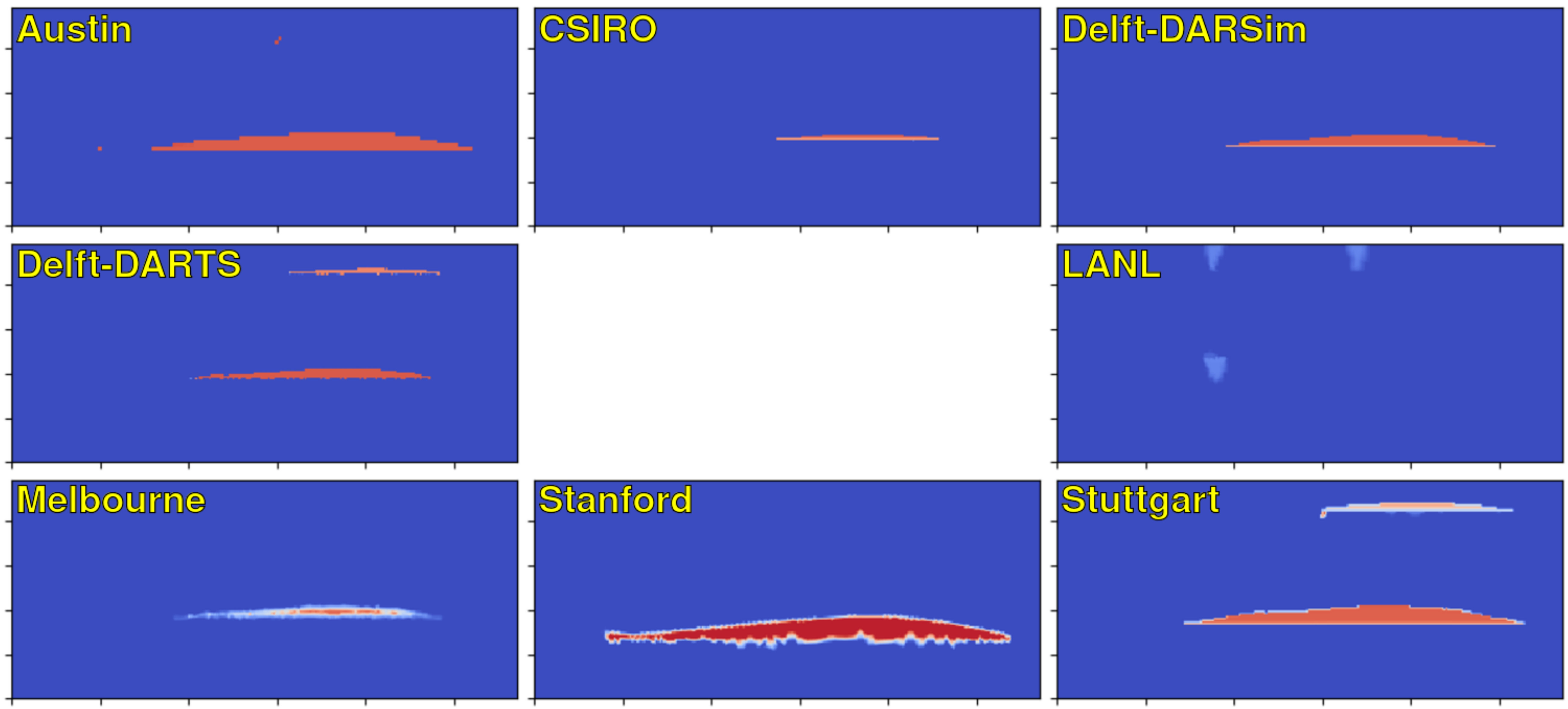}
\caption{Spatial distribution of gaseous \cotwo after \qty{72}{\hour}. The minimum for the color map is at 0 \cotwo saturation indicated by blue, the maximum at 1 indicated by red.}
\label{fig:saturation_three}
\end{figure}

\begin{figure}[hbt]
\centering
\includegraphics[width=0.99\textwidth]{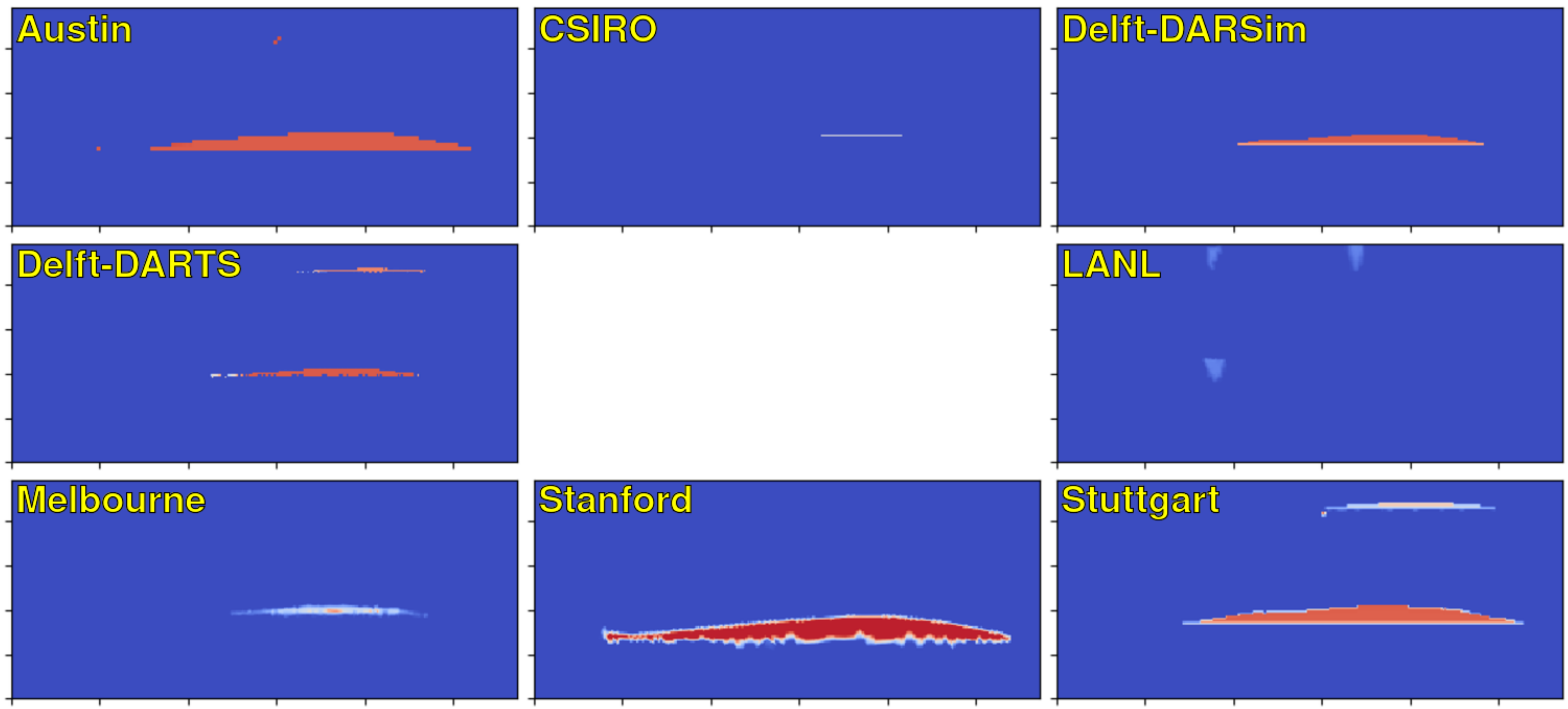}
\caption{Spatial distribution of gaseous \cotwo after \qty{96}{\hour}. The minimum for the color map is at 0 \cotwo saturation indicated by blue, the maximum at 1 indicated by red.}
\label{fig:saturation_four}
\end{figure}

\begin{figure}[hbt]
\centering
\includegraphics[width=0.99\textwidth]{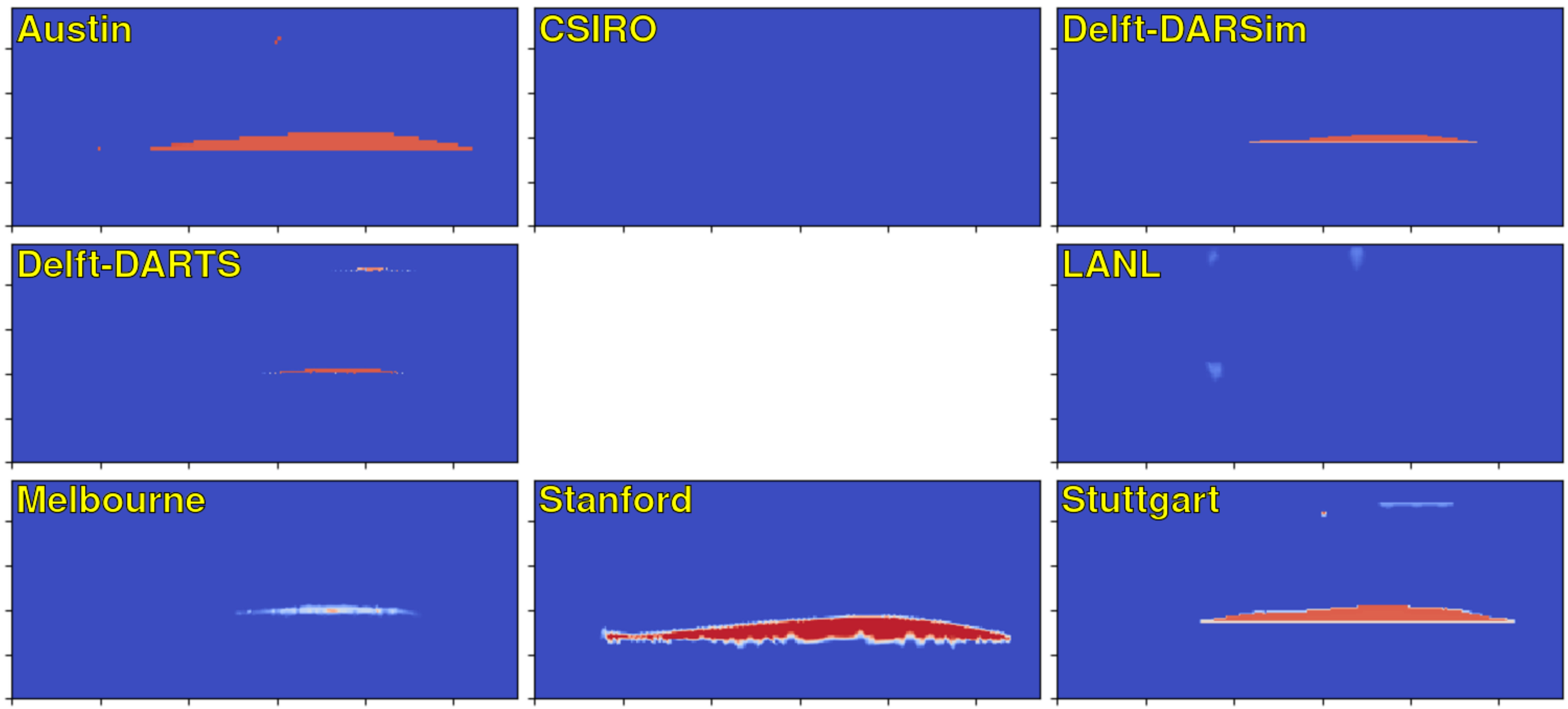}
\caption{Spatial distribution of gaseous \cotwo after \qty{120}{\hour}. The minimum for the color map is at 0 \cotwo saturation indicated by blue, the maximum at 1 indicated by red.}
\label{fig:saturation_five}
\end{figure}

Figures \ref{fig:saturation_one} to \ref{fig:saturation_five} visualize the reported saturation values for all participating groups at the selected daily time steps. Focusing first on Figure \ref{fig:saturation_one}, it can be observed that most participants report a very similar \cotwo plume shape under the lower fine sand barrier after \qty{24}{\hour}.
Moreover, no or almost no gaseous \cotwo is reported within Box B after one day. Considerably less agreement can be seen for the upper barrier in the right part of the domain. This can be explained by the fact that the amount of \cotwo injected in the lower and upper part differs by a factor of more than 2 and, correspondingly, a variation in the dissolution behavior becomes visible earlier in the upper part of the domain.
With \hw and \lanl, two participants report that no or almost no gaseous \cotwo is present throughout the domain after the first day of simulation. In case of \hw, this is due to the choice of the van-Genuchten relationship for the capillary pressure, as explained in more detail below in Section \ref{sec:results_time_series}. The reported results are the ones with the smallest capillary fringe that was possible to resolve within the computing time constraints and an overestimation of dissolution was anticipated. The situation is different for \lanl, where \cotwo leaves the system because almost no trapping occurs, see also below.

Examining the saturation distributions over the different time steps in Figures \ref{fig:saturation_one} to \ref{fig:saturation_five} reveals the effect of the varying dissolution behaviors. 
In particular, \csiro, \darsim, \darts and \melbourne report a vanishing \cotwo gas plume over time, while the plume shape stays rather constant for \austin, \stanford and \stuttgart. Starting with \qty{72}{\hour}, \hw did not report any spatial map data.

\subsubsection{Concentration}

\begin{figure}[hbt]
\centering
\includegraphics[width=0.99\textwidth]{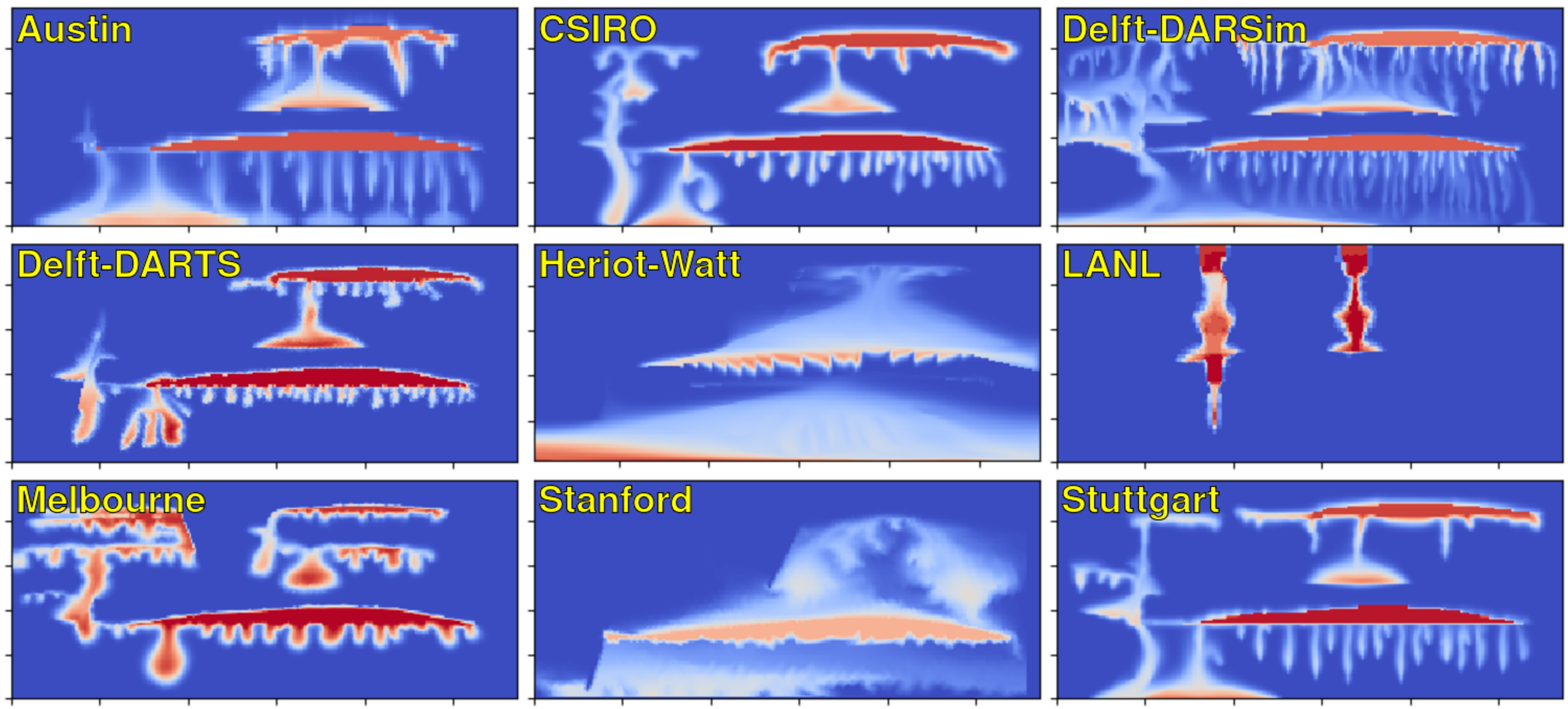}
\caption{Spatial distribution of \cotwo concentration in the liquid phase after \qty{24}{\hour}. The minimum for the color map is at \qty{0}{\kilogram\per\cubic\meter} indicated by blue, the maximum at \qty{1.8}{\kilogram\per\cubic\meter} indicated by red.}
\label{fig:concentration_one}
\end{figure}

\begin{figure}[hbt]
\centering
\includegraphics[width=0.99\textwidth]{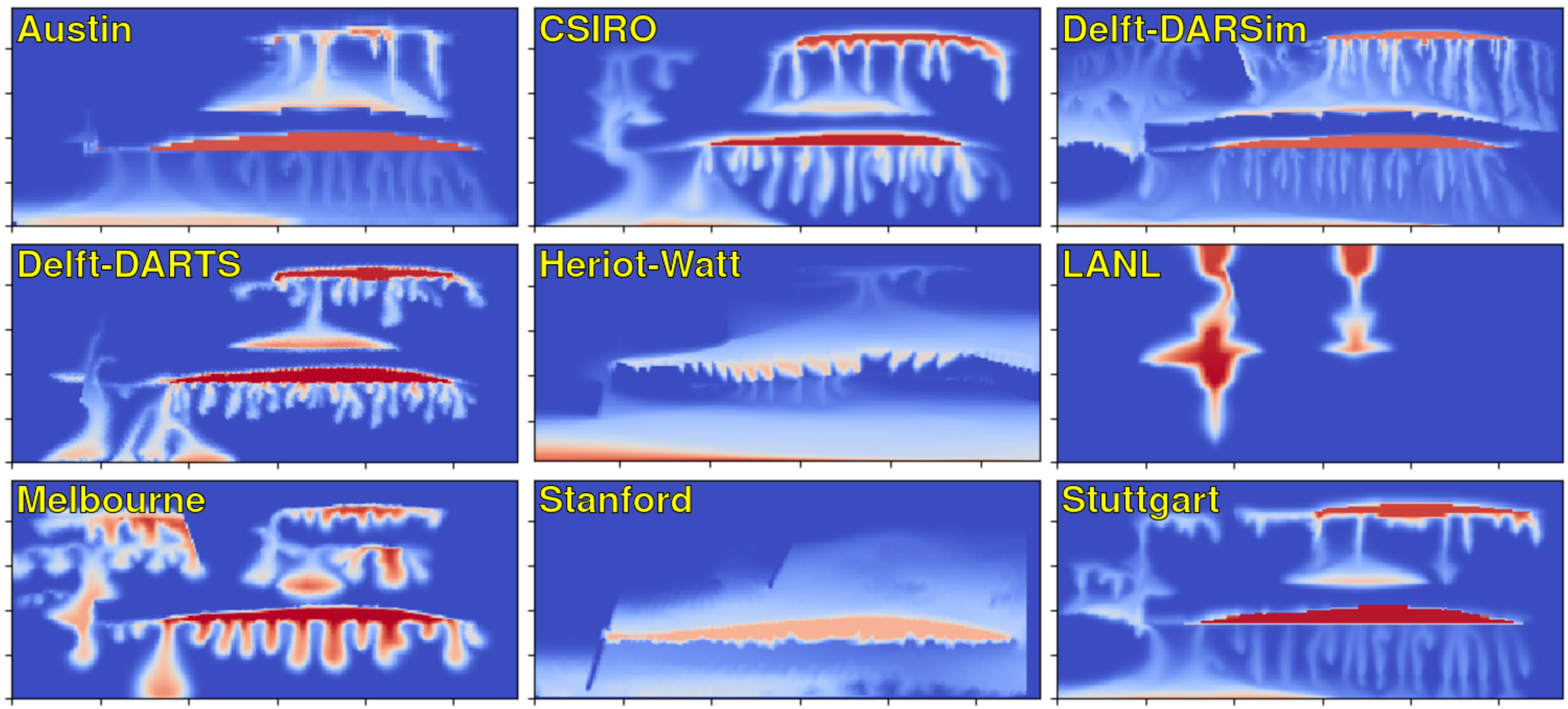}
\caption{Spatial distribution of \cotwo concentration in the liquid phase after \qty{48}{\hour}. The minimum for the color map is at \qty{0}{\kilogram\per\cubic\meter} indicated by blue, the maximum at \qty{1.8}{\kilogram\per\cubic\meter} indicated by red.}
\label{fig:concentration_two}
\end{figure}

\begin{figure}[hbt]
\centering
\includegraphics[width=0.99\textwidth]{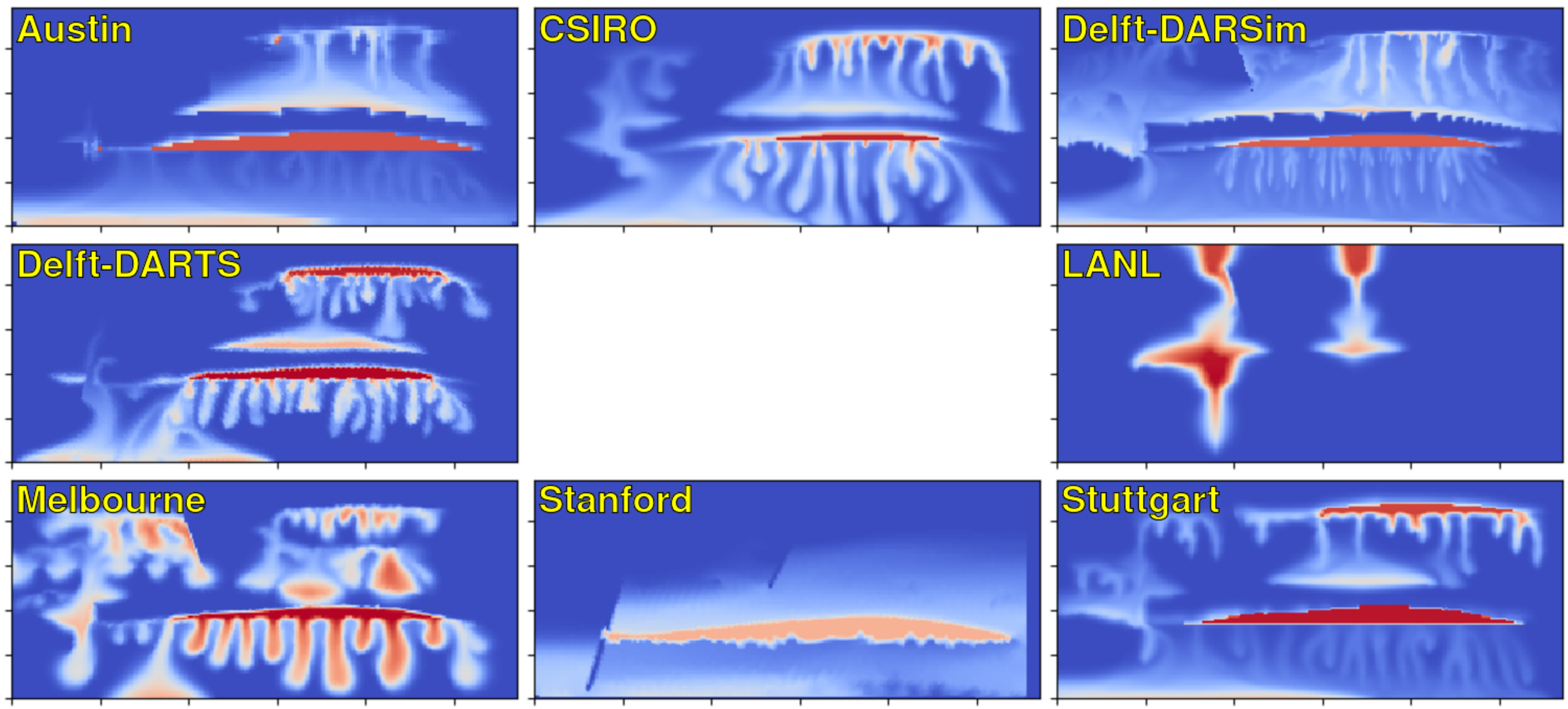}
\caption{Spatial distribution of \cotwo concentration in the liquid phase after \qty{72}{\hour}. The minimum for the color map is at \qty{0}{\kilogram\per\cubic\meter} indicated by blue, the maximum at \qty{1.8}{\kilogram\per\cubic\meter} indicated by red.}
\label{fig:concentration_three}
\end{figure}

\begin{figure}[hbt]
\centering
\includegraphics[width=0.99\textwidth]{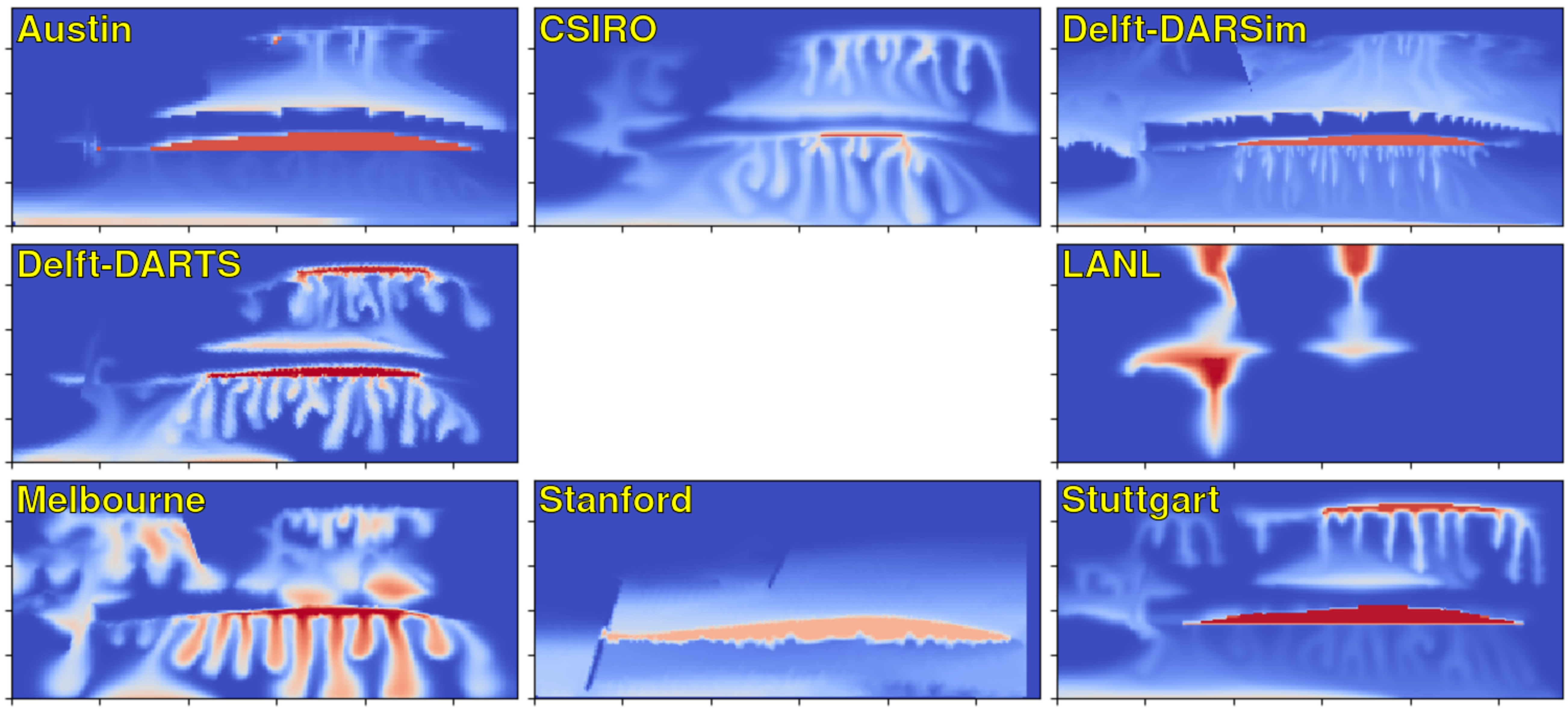}
\caption{Spatial distribution of \cotwo concentration in the liquid phase after \qty{96}{\hour}. The minimum for the color map is at \qty{0}{\kilogram\per\cubic\meter} indicated by blue, the maximum at \qty{1.8}{\kilogram\per\cubic\meter} indicated by red.}
\label{fig:concentration_four}
\end{figure}

\begin{figure}[hbt]
\centering
\includegraphics[width=0.99\textwidth]{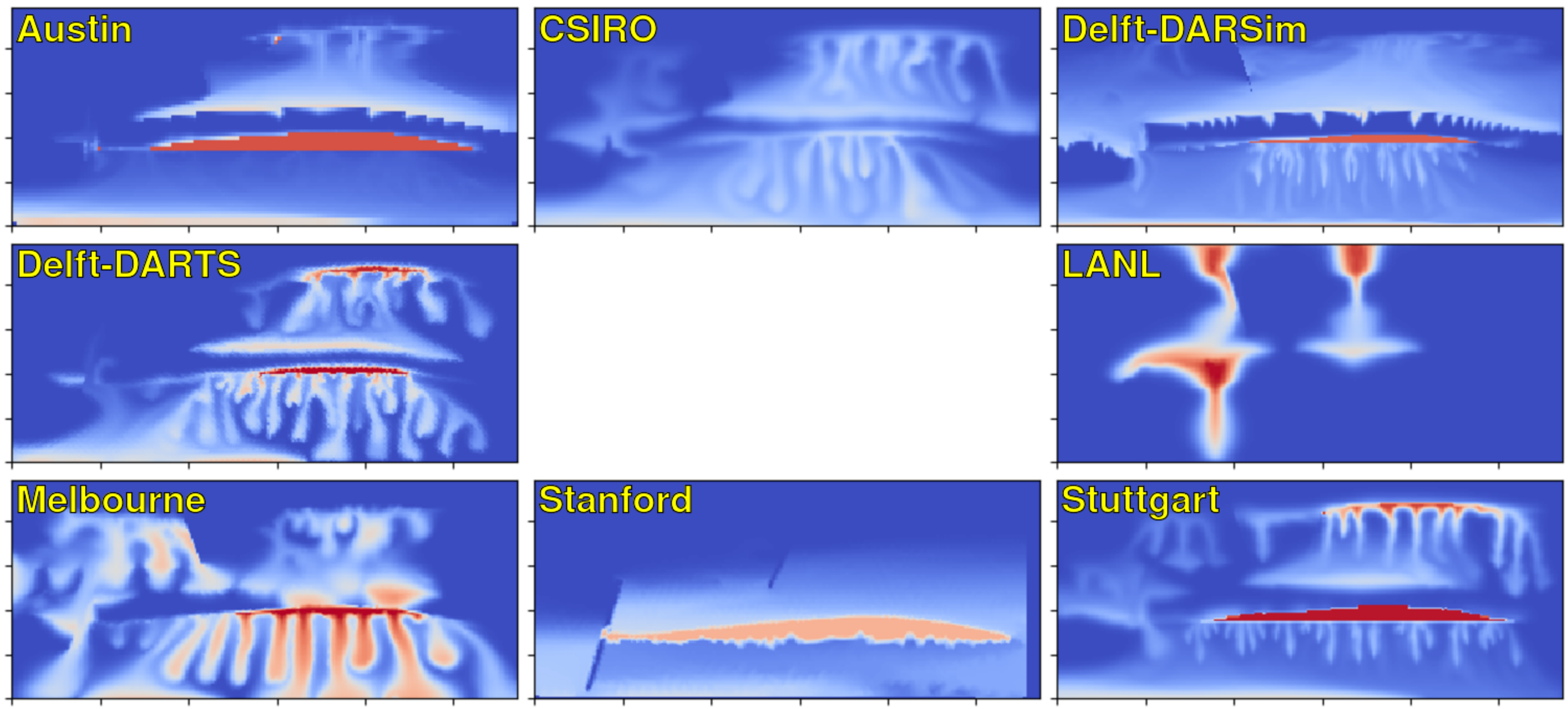}
\caption{Spatial distribution of \cotwo concentration in the liquid phase after \qty{120}{\hour}. The minimum for the color map is at \qty{0}{\kilogram\per\cubic\meter} indicated by blue, the maximum at \qty{1.8}{\kilogram\per\cubic\meter} indicated by red.}
\label{fig:concentration_five}
\end{figure}

Analogous to the saturation, Figures \ref{fig:concentration_one} to \ref{fig:concentration_five} visualize the reported concentration values for all participating groups at the selected daily time steps.
While at first glance, the variation in the results appears to be larger than for the saturation, the reported qualitative behavior is similar for most groups.
The \cotwo dissolves into the liquid phase and, due to the density difference between pure and \cotwo-enriched water, the latter is moving downwards by developing fingers. This motion is impeded by fine-sand barriers or the bottom of the domain.
A clear outlier to this rather uniform qualitative behavior is given by \lanl, whose simulations indicate that \cotwo has moved relatively straight upward without being hindered substantially by the fine-sand barriers. A variety of possible reasons exist, ranging from differently interpreted facies geometries and realized computational grids over too small variations in spatial parameters up to insufficient constitutive relationships. As running two codes with PFLOTRAN and FEHM yielded similar results, the exact reason could not be determined during the course of the study.

The main quantitative differences which can be observed among the remaining groups arise due to the different speeds at which dissolution is taking place. In particular, dissolution for \hw and \stanford appears to be much faster than for the other participating groups.
Moreover, quite some disagreement can be observed on how much \cotwo is reaching the upper left part of the domain, i.e., Box B, via the corresponding fault zone.

Another interesting measure is the amount and respective thickness in horizontal direction of the evolving fingers. Differences here can be largely attributed to correspondingly different grid resolutions. For example, the participating groups \csiro, \darsim and \darts with relatively high resolution (cf. Table \ref{tab:choices}) show substantially more and thinner fingers than \austin and \melbourne with a relatively low resolution. 

\subsubsection{Quantitative Comparison}
\label{sec:wasserstein_models}
As a quantitative measure, we apply the Wasserstein metric to analyze the difference between two snapshots. This metric works on distributions of equal mass and measures ``the minimal effort required to reconfigure the mass of one distribution in order to recover the other distribution''~\cite{Panaretos:2019:SAW}
In order to apply the Wasserstein metric to the reported results, which in general have a slightly different mass (see detailed discussion in section \ref{sec:results_time_series_mass}), we first approximate roughly the \cotwo mass density in each cell by combining the reported concentration and saturation values via the formula
\[
\widetilde{m} = \varrho_\text{g}s + c(1-s).
\]
Above, $s$ and $c$ indicate the saturation and concentration value, while the density $\varrho_\text{g}$ of gaseous \cotwo  is set to \qty{2}{\kg\per\cubic\meter} to reflect the experimental conditions. The resulting values can be visualized by corresponding grayscale pictures which have been uploaded to the participants' data repositories. The final step to make these values comparable is their normalization such that they can be treated formally as two-dimensional probability distributions over the experimental domain. Given the normalized values, the Python library POT~\cite{Flamary:2021:POT} can be applied to calculate the Wasserstein distances. The values are listed in Appendix \ref{sec:distances} for every requested individual timestep. The full data including distances between results from different timesteps is provided in the FluidFlower general GitHub repository. This approach  provides a reasonable approximation for the groups with approximately equal mass in the reported results, however, it is not appropriate for the results from \lanl, whose simulations indicate that a significant fraction of the injected mass leaves the domain. Therefore, the results from \lanl are excluded from the Wasserstein distance calculations.  

\begin{figure}[hbt]
\centering
\includegraphics[width=0.99\textwidth]{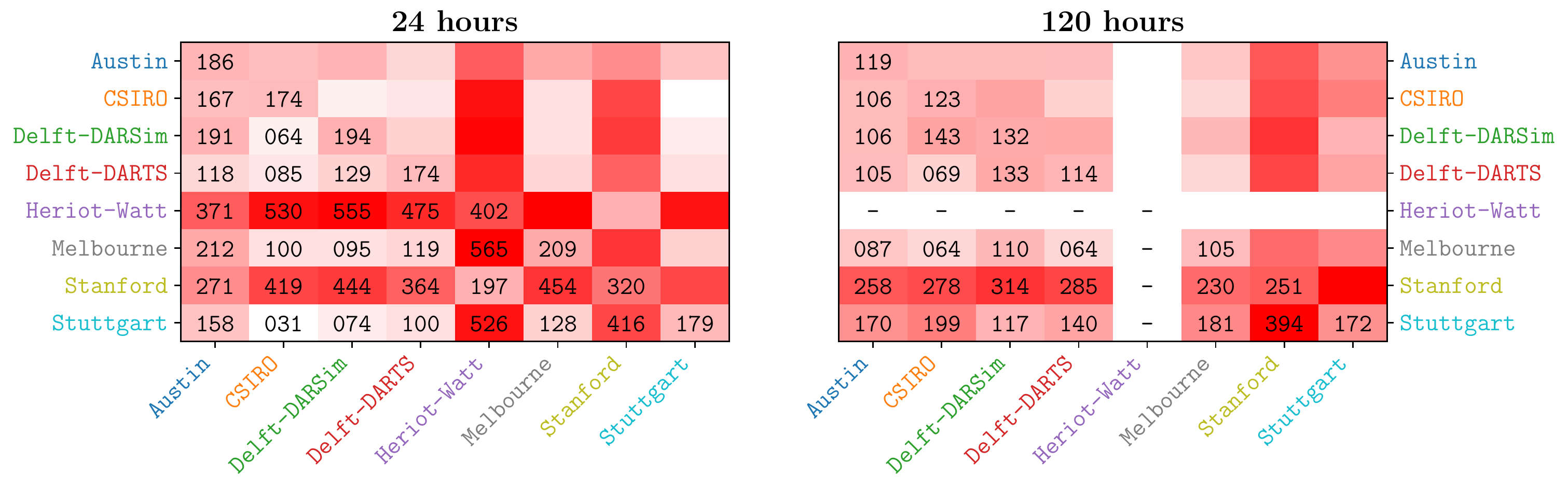}
\caption{Wasserstein distances in gram times centimeter for the first and last time step. Colors range from white for low values to red for high values. A value on a diagonal is the mean value of the respective row/column. Values above the diagonal are not displayed as they are symmetric. As no spatial map has been reported by \hw for \qty{120}{\hour}, the corresponding fields are left empty.}
\label{fig:pcolor}
\end{figure}

We illustrate the calculated Wasserstein distances exemplarily for the first and last time step in Figure \ref{fig:pcolor}.
The values have been dimensionalized by multiplying with the real mass of \cotwo in the system and are provided in units of gram times centimeter. Thus a value of 100 gr.cm corresponds to one gram of mass (e.g. about 20\% of the \cotwo in the system) being shifted by one meter (e.g. about one third of the full simulation domain). Values on the order of 100 gr.cm or less thus correspond to what we consider relatively close results, while results in significant excess of 100 gr.cm indicate substantial discrepancies. Figure \ref{fig:pcolor} thus quantifies the qualitative results discussed in the subsections above. In particular, the spatial maps from \hw and \stanford show the largest distances to the other groups over all time steps. Their mean distances are between two and three times larger than the ones from the other groups, due to their different dissolution behavior.
Overall, the mean distances are mostly decreasing from the first to the last time step, as \cotwo further dissolves in the water and its mass distributes more over the domain.

\subsection{Dense data time series}
\label{sec:results_time_series}

The participating groups were instructed to report several scalar SRQs in ten-minute intervals over a time span of five days: total mass of \cotwo inside the domain, pressure at two locations, phase composition in Boxes A and B, as well as convection in Box C.

\subsubsection{Total mass of \cotwo}
\label{sec:results_time_series_mass}

\begin{figure}[hbt]
\centering
\includegraphics[width=0.8\textwidth]{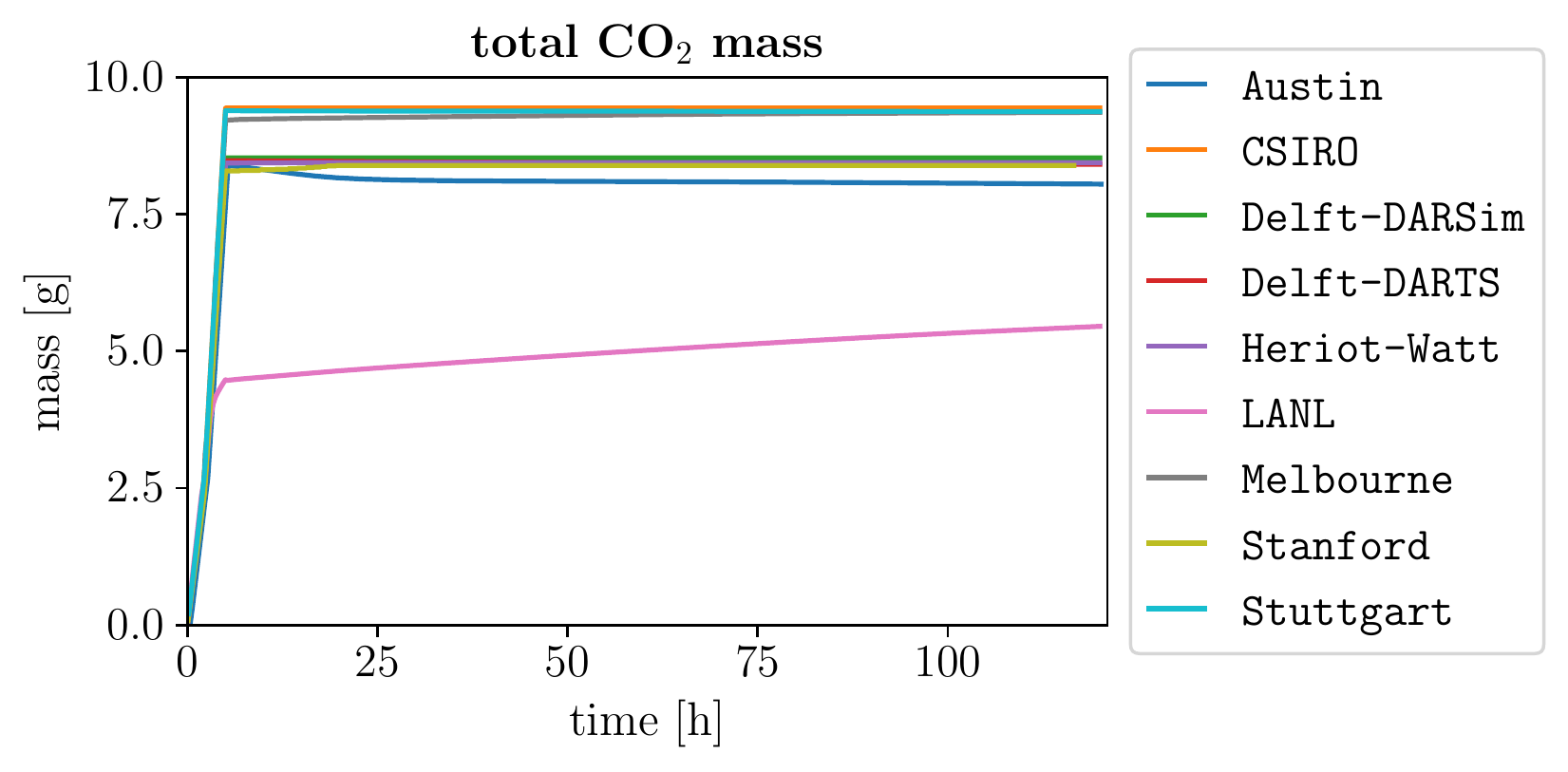}
\caption{Temporal evolution of the total \cotwo mass inside the computational domain.}
\label{fig:time_series_co2mass}
\end{figure}

Figure \ref{fig:time_series_co2mass} depicts the temporal evolution of the total mass of \cotwo inside the computational domain, as reported by the different participating groups. 
The benchmark description prescribes the injection rates in terms of Standard Cubic Centimeters per Minute (SCCM) \cite{Nordbotten:2022:FBD}. While the underlying standard conditions are not explicitly specified, the instrument employed by \expub uses the NIST definition of standard conditions, i.e.  \qty{293.15}{\kelvin} and \qty{1.013}{\bar}. This would yield a final total mass of approximately \qty{8.5}{\gram}, assuming that no \cotwo leaves the domain. While the majority of the modeling groups employed the corresponding interpretation of standard conditions, three groups report a higher value of approximately \qty{9.4}{\gram}.
With \lanl, one group reports considerable lower values which is due to the fact that \cotwo leaves the domain, as has been explained in more detail in Section \ref{sec:results_spatial_maps}. In most results, the total amount of \cotwo stays constant after injection stops, indicating that no mass leaves the system. Nevertheless, some groups report a further increase
or also a further decrease,
which can partially be explained by numerical effects~\cite{Youssef:2023:SAS} or again the circumstance that gaseous \cotwo leaves the computational domain, respectively. 

\subsubsection{Pressure}
\label{sec:time_series_pressure}

\begin{figure}[hbt]
\centering
\includegraphics[width=0.95\textwidth]{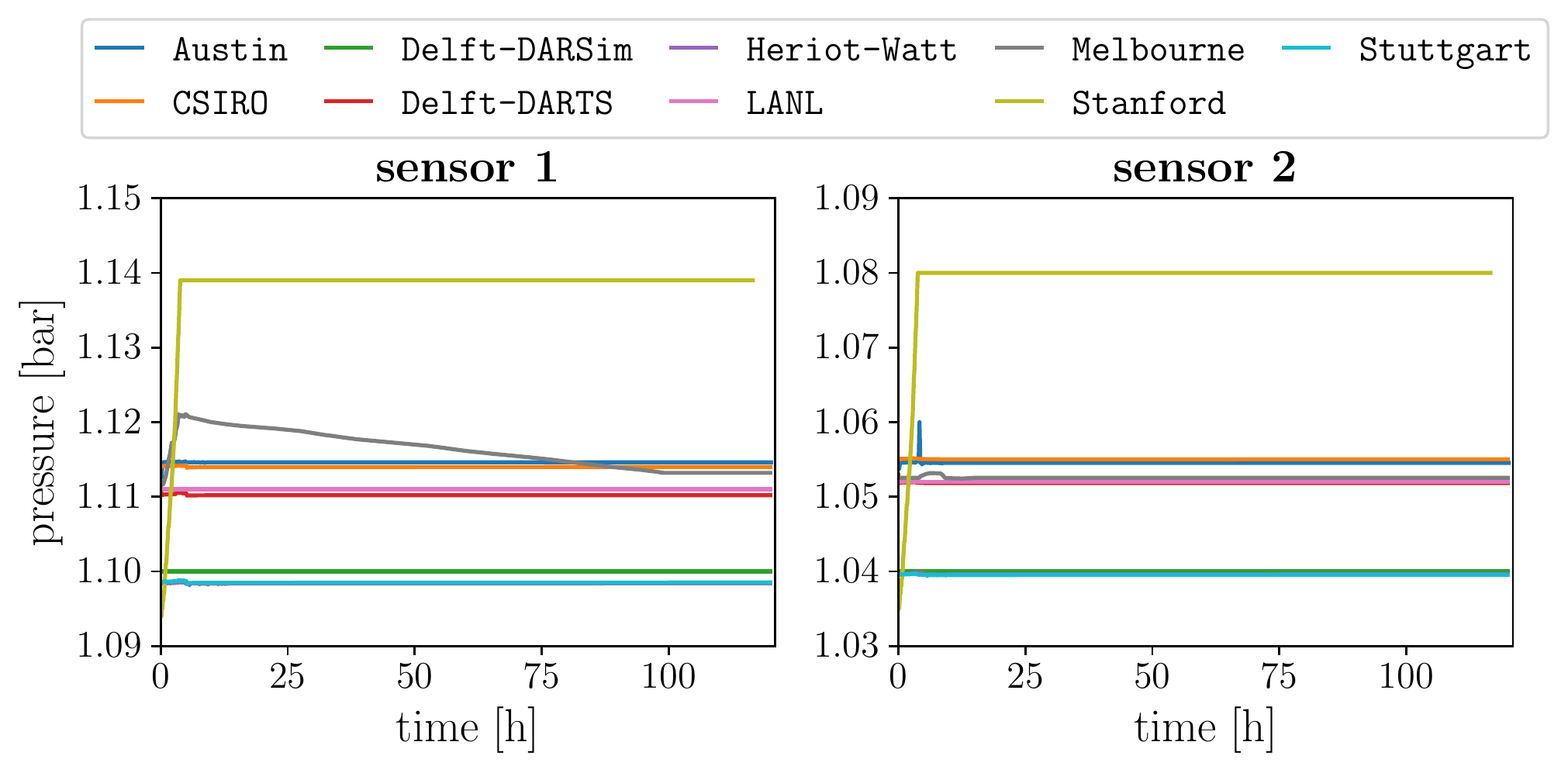}
\caption{Temporal evolution of the pressure at two locations inside the computational domain, Sensor 1 (left) and Sensor 2 (right).}
\label{fig:time_series_pressure}
\end{figure}

\begin{figure}[hbt]
\centering
\includegraphics[width=0.95\textwidth]{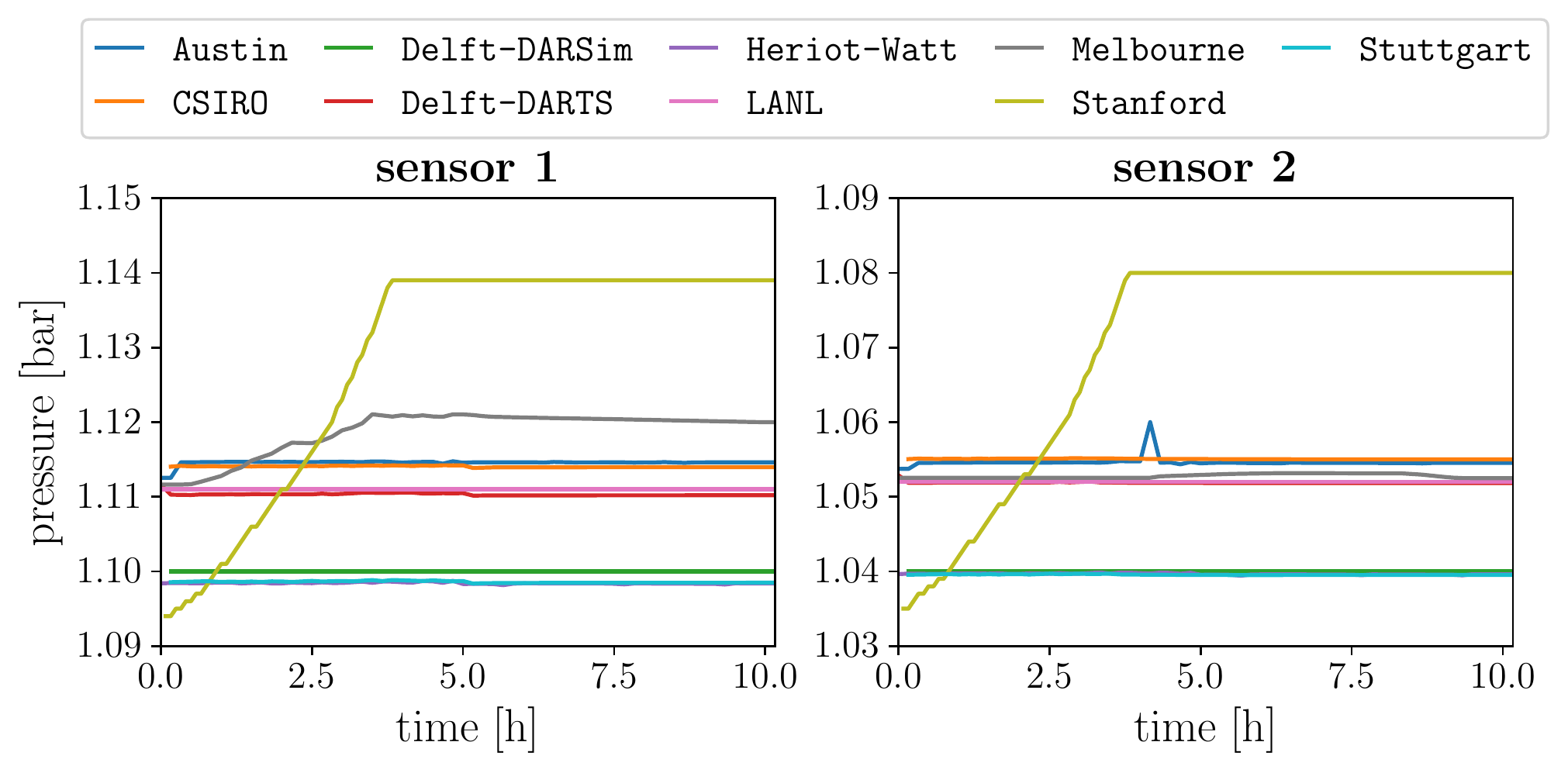}
\caption{Temporal evolution of the pressure at two locations inside the computational domain, Sensor 1 (left) and Sensor 2 (right). Zoom into the first six hours.}
\label{fig:time_series_pressure_zoom}
\end{figure}

The next reported SRQ is the temporal evolution of the pressure, measured at two sensors in the domain. Figure \ref{fig:time_series_pressure} illustrates the reported results.
Most of the results show at most a minor influence of the \cotwo injection on the observed pressure values. The pressure at each sensor stays rather constant at the prescribed initial and possibly boundary conditions which correspond to an assumed ambient atmospheric pressure plus the effect of the water table. Nevertheless, two groups, \stanford and \melbourne, report a considerable influence of the injection processes. In order to examine this in more detail, Figure \ref{fig:time_series_pressure_zoom} depicts a zoom into the first ten hours of simulation.
The results from \melbourne show a considerable increase only for the first sensor which decays slowly to a constant level after the stop of injection. Here, the difference in the buildup between the two sensors can be explained by their respective proximity to the injection wells.
In contrast to this, \stanford reports the same pressure buildup for both sensors. This can be explained by the fact that no-flow boundary conditions are used everywhere and the fluids are assumed to be incompressible. Notably, both groups report a stop of the pressure buildup at around 3.5 hours, before the stop of \cotwo injection at 5 hours.

\subsubsection{Phase composition}
\label{sec:time_series_composition}

\begin{figure}[hbt]
\centering
\includegraphics[width=0.95\textwidth]{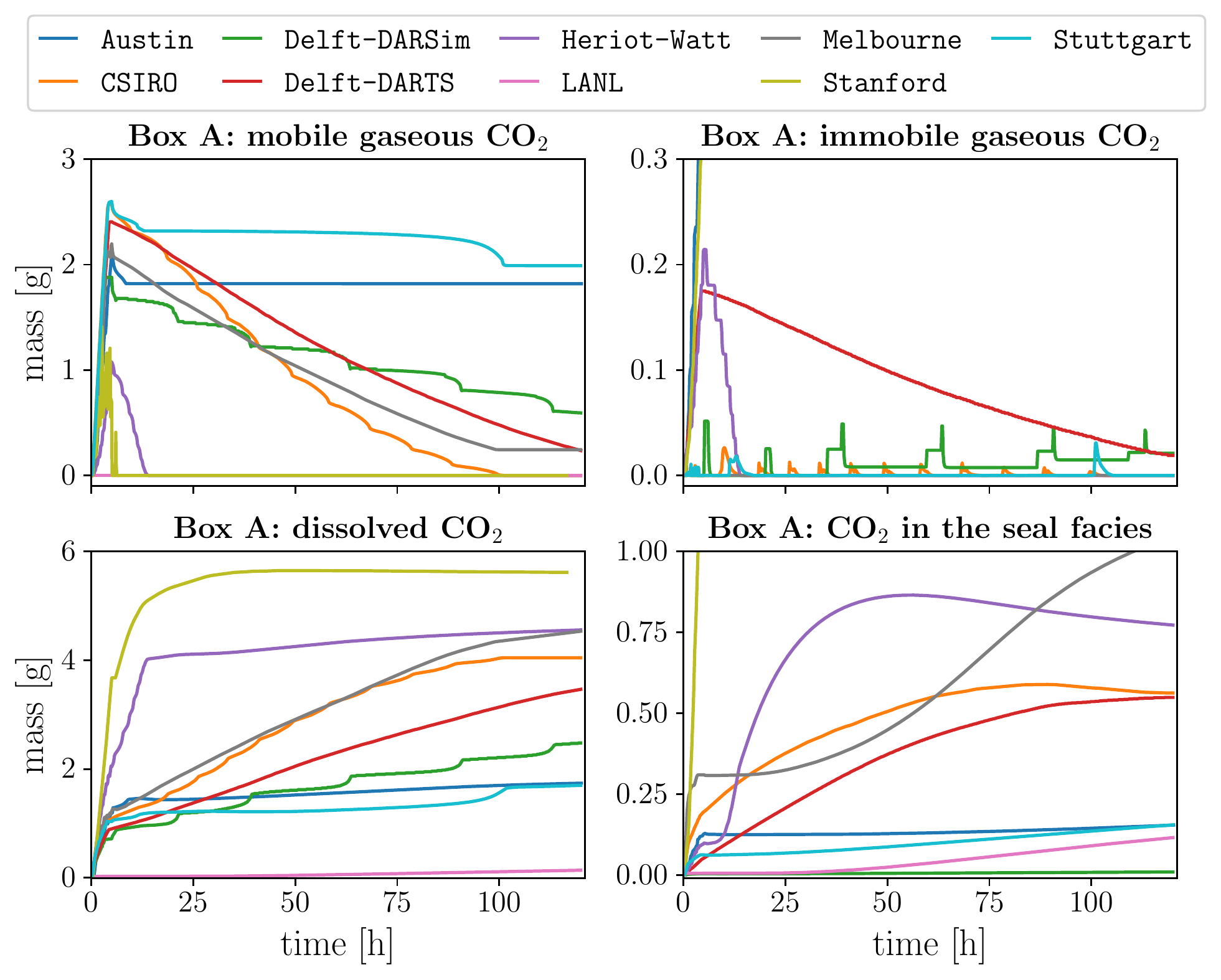}
\caption{Temporal evolution of the \cotwo phase distribution in Box A.}
\label{fig:time_series_boxA}
\end{figure}

In the following, we discuss the reported distribution of \cotwo over the two fluid phases in Boxes A and B. In particular, the participants reported the evolution of the amount of mobile and immobile gaseous \cotwo, \cotwo dissolved in the liquid phase, as well as \cotwo contained in the seal facies. We first focus on Box A and the respective Figure \ref{fig:time_series_boxA}.
It can be seen immediately that the variation of the results across the participating groups is much larger than for the previous SRQs. All results have in common that mobile gaseous \cotwo reaches a peak value at approximately five hours (coinciding with the injection stop) and then dissolves at different rates. Eight results can be grouped into three clusters showing a similar rate. The largest cluster consists of the participants \csiro, \darsim, \darts and \melbourne. Here, the dissolution takes place over the whole simulation period at an intermediate rate compared to the other two clusters. The two participants \austin and \stuttgart both show after an initial decay a very slow dissolution behavior. In contrast to this, \hw and \stanford predict the fastet dissolution with zero mobile gaseous \cotwo left after less than one day. However, \stanford reports a very high amount of gaseous \cotwo becoming immobile, due to their employed identification of immobile gas leading to an overestimation. An outlier can be identified by \lanl, where no \cotwo at all reaches Box A. All these observations are consistent with the results and discussion concerning the spatial maps in Section \ref{sec:results_spatial_maps}. In addition here, a remarkable characteristic is the step-like progression of several curves, as reported particularly by \csiro, \darsim and \stuttgart. This numerical effect is due to grid-dependent bursts in dissolution when the water-gas contact coincides with cell faces. It has also been observed initially by \hw, who decided to employ the capillary pressure - saturation relationship by van Genuchten for the coarser sands in order to prevent the effect, see also Table \ref{tab:choices}.

\begin{figure}[hbt]
\centering
\includegraphics[width=0.95\textwidth]{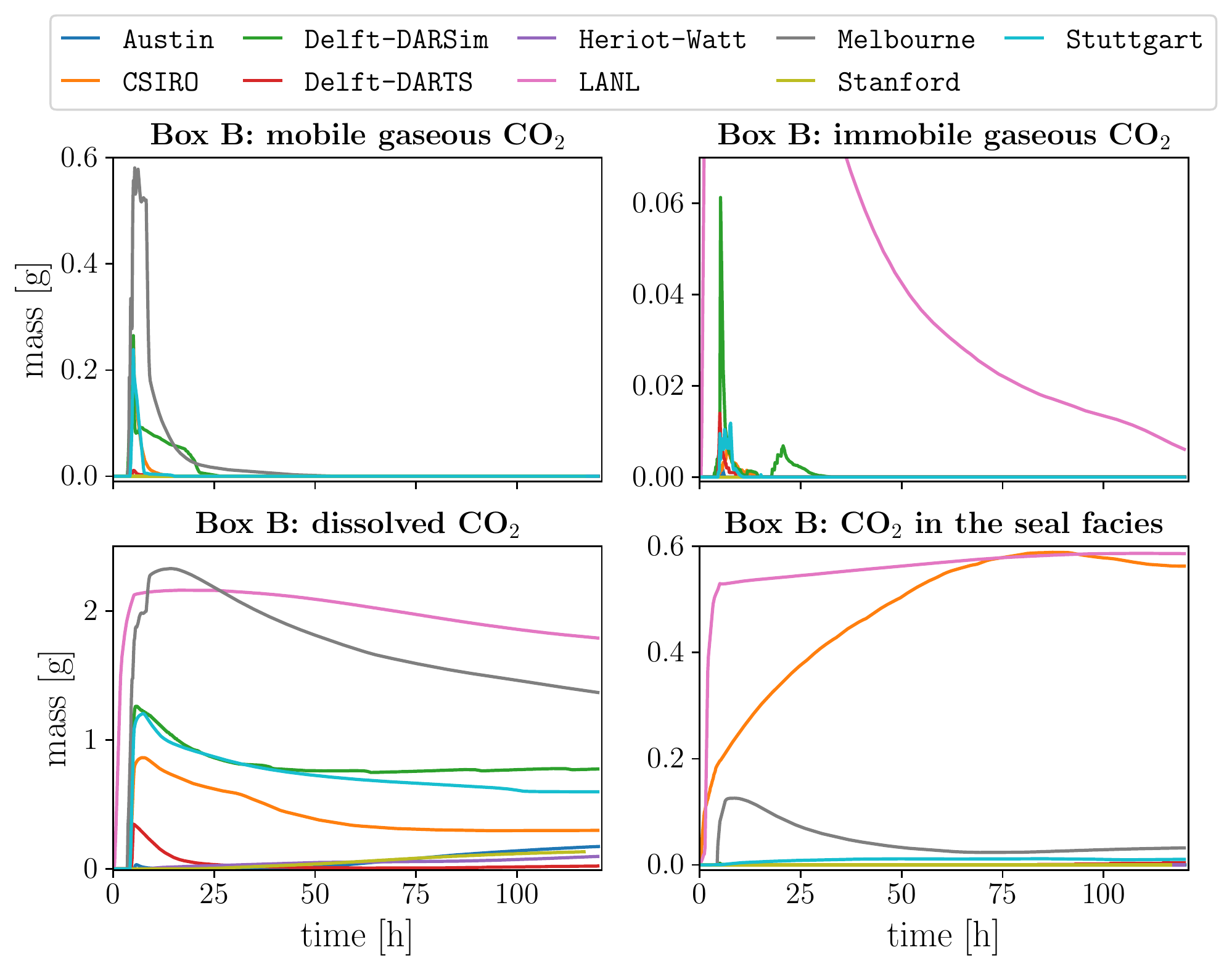}
\caption{Temporal evolution of the \cotwo phase distribution in Box B.}
\label{fig:time_series_boxB}
\end{figure}

Turning to Box B and Figure \ref{fig:time_series_boxB}, the results exhibit even more variation. This can be attributed to the location of the box with the challenge of quantifying how much \cotwo reaches the fault zone in the lower left and subsequently the upper left region of the domain.
While all participants predict the disappearance of mobile gaseous \cotwo after at most two days, the peak amount varies strongly between zero and \qty{0.6}{\gram}. These different peak amounts together with different dissolution rates explain the high variation in dissolved \cotwo as seen in Figure \ref{fig:time_series_boxB}. On a positive note, almost all models predicting a substantial amount of \cotwo in Box B report very similar times of appearance.

\subsubsection{Convection}

\begin{figure}[hbt]
\centering
\includegraphics[width=0.8\textwidth]{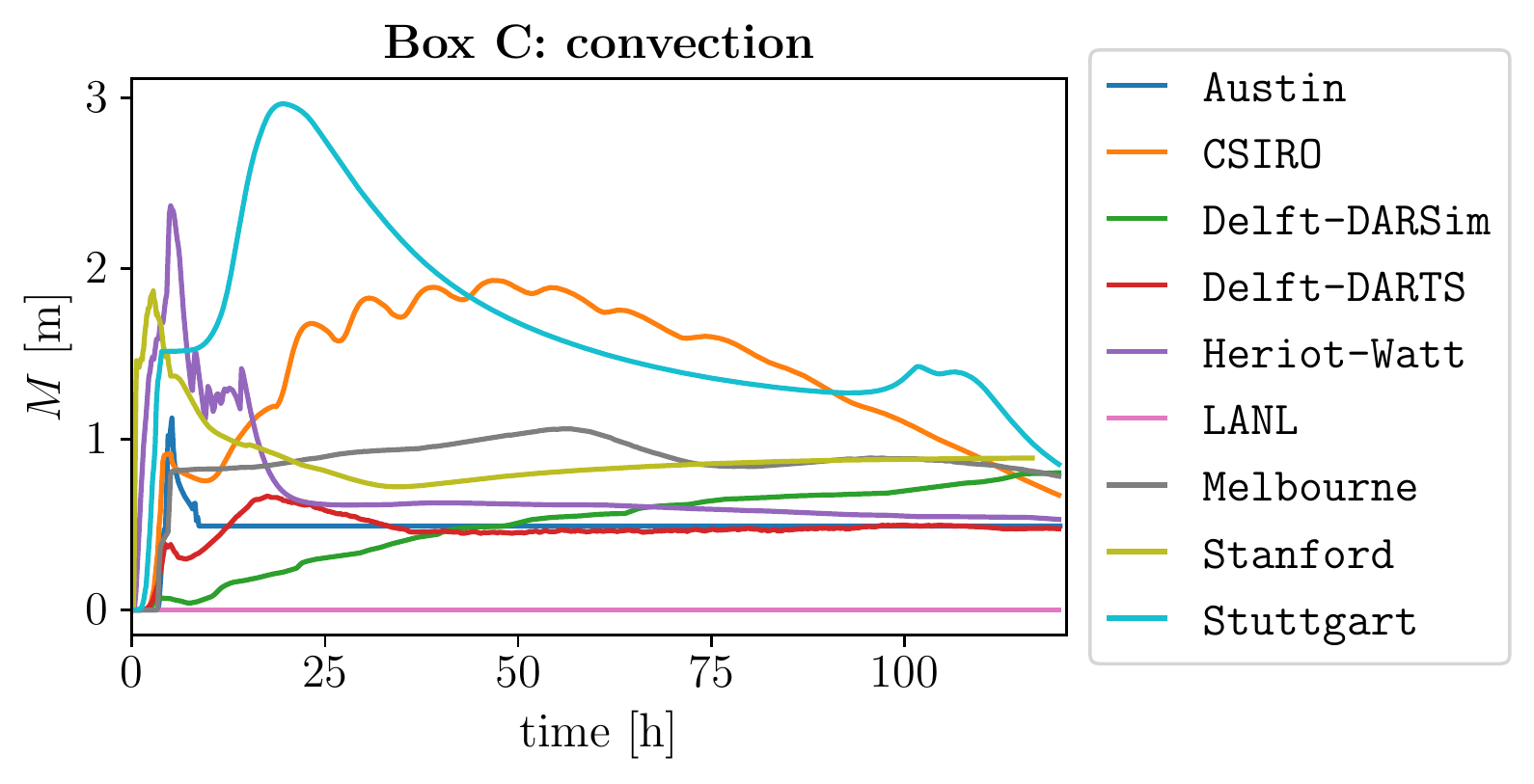}
\caption{Temporal evolution of $M(t)$ as a measure for convection in Box C.}
\label{fig:time_series_boxC}
\end{figure}

As a measure for convection, the participants where asked to report the total variation of concentration within Box C over time, see the definition of $M(t)$ in \cite[Section 2.8.3]{Nordbotten:2022:FBD}.
The results are depicted in Figure \ref{fig:time_series_boxC}.
A relatively large spread with peak values ranging from 0 to \qty{3}{\meter} can be observed. Also the dynamic behavior is very different, ranging from a monotone increase to rather strong oscillations. Nevertheless, most participants report a stabilization over time to a stationary value between 0.5 and \qty{1}{\meter}.

\subsection{Sparse data}
\label{sec:results_sparse_data}

In this section, we describe the reported so-called sparse data.
Each of the sparse data items had to be reported as six numbers, representing the prediction of the mean quantity as obtained by the experiments (stated in terms of P10, P50 and P90 values), as well as the prediction in the standard deviation of the quantity over the ensemble of experiments (again stated as P10, P50, and P90 values). Since most groups didn't report any P10 and P90 values for the expected standard deviations, we only consider the P50 values for the following comparisons. As basis for generating the predictions and uncertainties, any preferred methodology could be chosen, ranging from ensemble runs and formal methods of uncertainty quantification until human intuition from experience. We start with the maximum pressure at the two sensors, then focus on the times of maximum mobile gaseous \cotwo in Box A and onset of convective mixing in Box C, before we investigate the predicted phase distributions after three days in Boxes A and B.

\subsubsection{Maximum pressure at the two sensors}
\label{sec:maxpress}

\begin{figure}[hbt]
\centering
\includegraphics[width=0.99\textwidth]{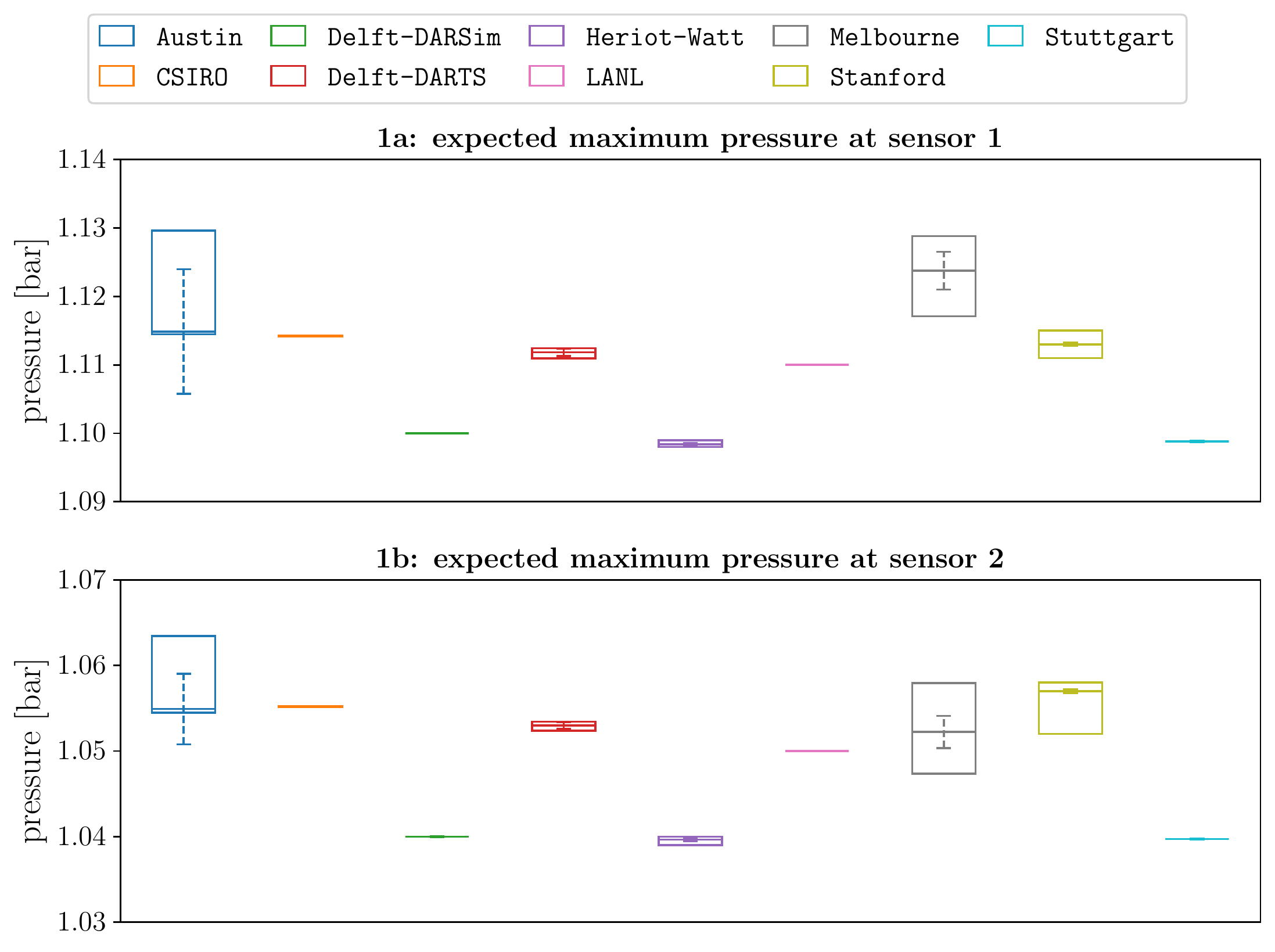}
\caption{Reported sparse data for the maximum pressure at sensors 1 and 2. Bottom, middle and top horizontal lines of the boxes indicate the reported P10, P50 and P90 values for the expected mean value, respectively. Dashed vertical lines extend from the mean values by $\pm$ the reported P50 of the expected standard deviations.}
\label{fig:sparse_pressure}
\end{figure}

The participants were asked for the expected maximum pressure at Sensors 1 and 2 as a proxy for assessing the risk of mechanical disturbance of the overburden. The reported values are depicted in Figure \ref{fig:sparse_pressure}.
As can be seen from the scaling of the vertical axis, all participating groups report very similar pressure values. Most groups also report P10, P50 and P90 values for the expected mean which are very close to each other, with the largest difference for one group being around \qty{10}{\milli\bar}. With \austin and \melbourne, only two groups expect any substantial standard deviation over the ensemble of experiments. The difference over all groups between the minimum P10 and maximum P90 reported pressure value is less than \qty{40}{\milli\bar} for each of the two sensors. This indicates that the typical variation in atmospheric pressure at the location of the experimental rig was not taken into account, exceeding \qty{50}{\milli\bar} over the winter months. Although the exact days of the experimental runs have not been provided explicitly to the participants, the information on the usual pressure variation is publicly available\footnote{\url{https://weatherspark.com/h/s/148035/2021/3/Historical-Weather-Winter-2021-at-Bergen-Flesland-Norway\#Figures-Pressure}}.

\subsubsection{Times of maximum mobile gaseous \cotwo in Box A and onset of convective mixing in Box C}

\begin{figure}[hbt]
\centering
\includegraphics[width=0.99\textwidth]{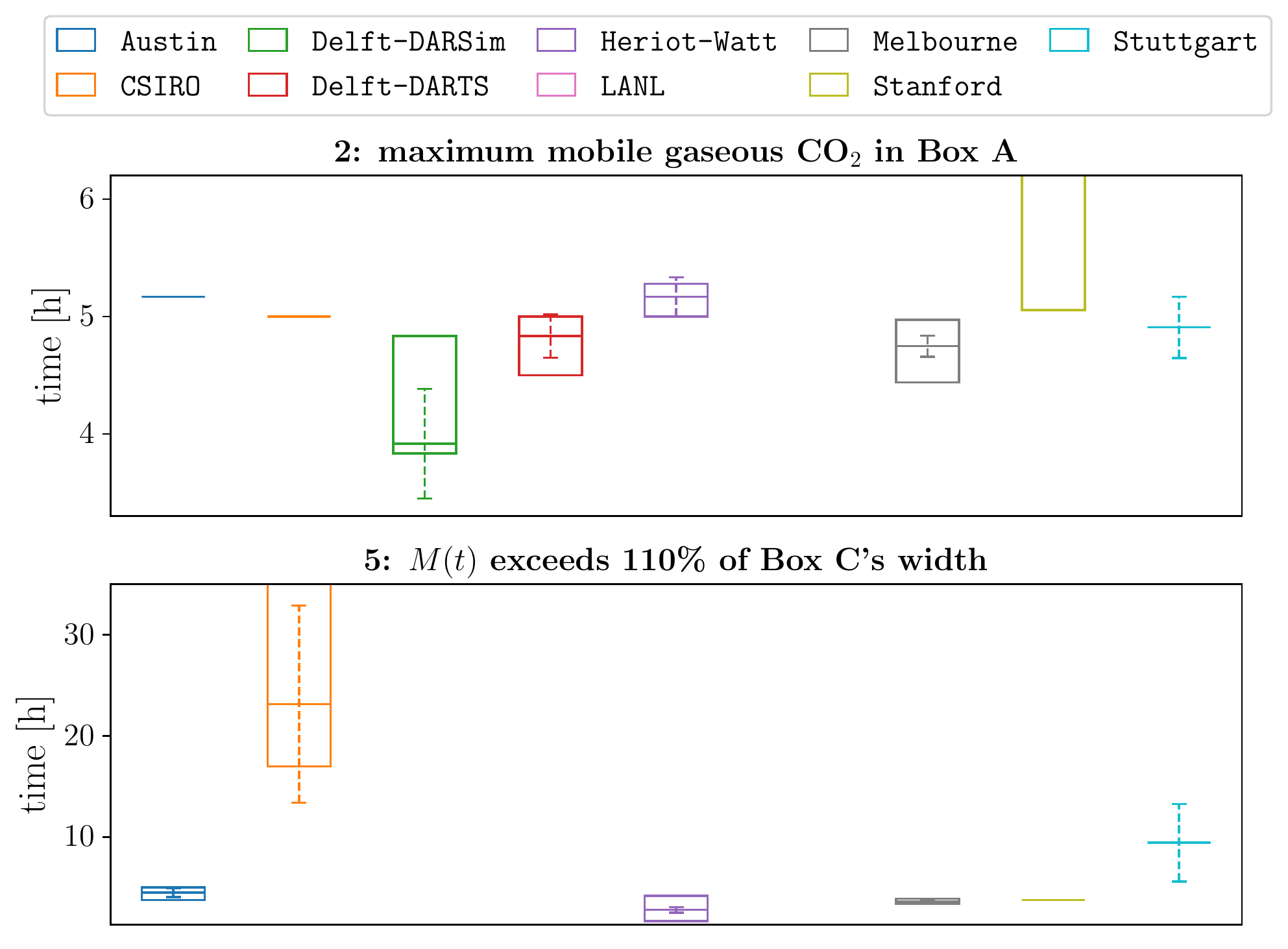}
\caption{Reported sparse data for the times of maximum mobile gaseous \cotwo in Box A (top) and for which the integral $M(t)$ first exceeds 110\% of the width of Box C (bottom). Bottom, middle and top horizontal lines of the boxes indicate the reported P10, P50 and P90 values for the expected mean value, respectively. Dashed vertical lines extend from the mean values by $\pm$ the reported P50 of the expected standard deviations.}
\label{fig:sparse_time}
\end{figure}

We now focus on the time of maximum mobile gaseous \cotwo in Box A as a proxy for when leakage risk starts declining. The corresponding reported values are visualized in the upper picture of Figure \ref{fig:sparse_time}.
The majority of the participating groups now report substantial differences between the P10 and P90 values of both the expected mean and standard deviation. Nevertheless, several groups are very certain on the expected mean value and report only very narrow ranges. The variation between the groups is considerably larger than for the pressure discussed above. This is well explainable by the larger variation in the modeling results as discussed in Sections \ref{sec:time_series_pressure} and \ref{sec:time_series_composition}.

\begin{figure}[hbt]
\centering
\includegraphics[width=0.8\textwidth]{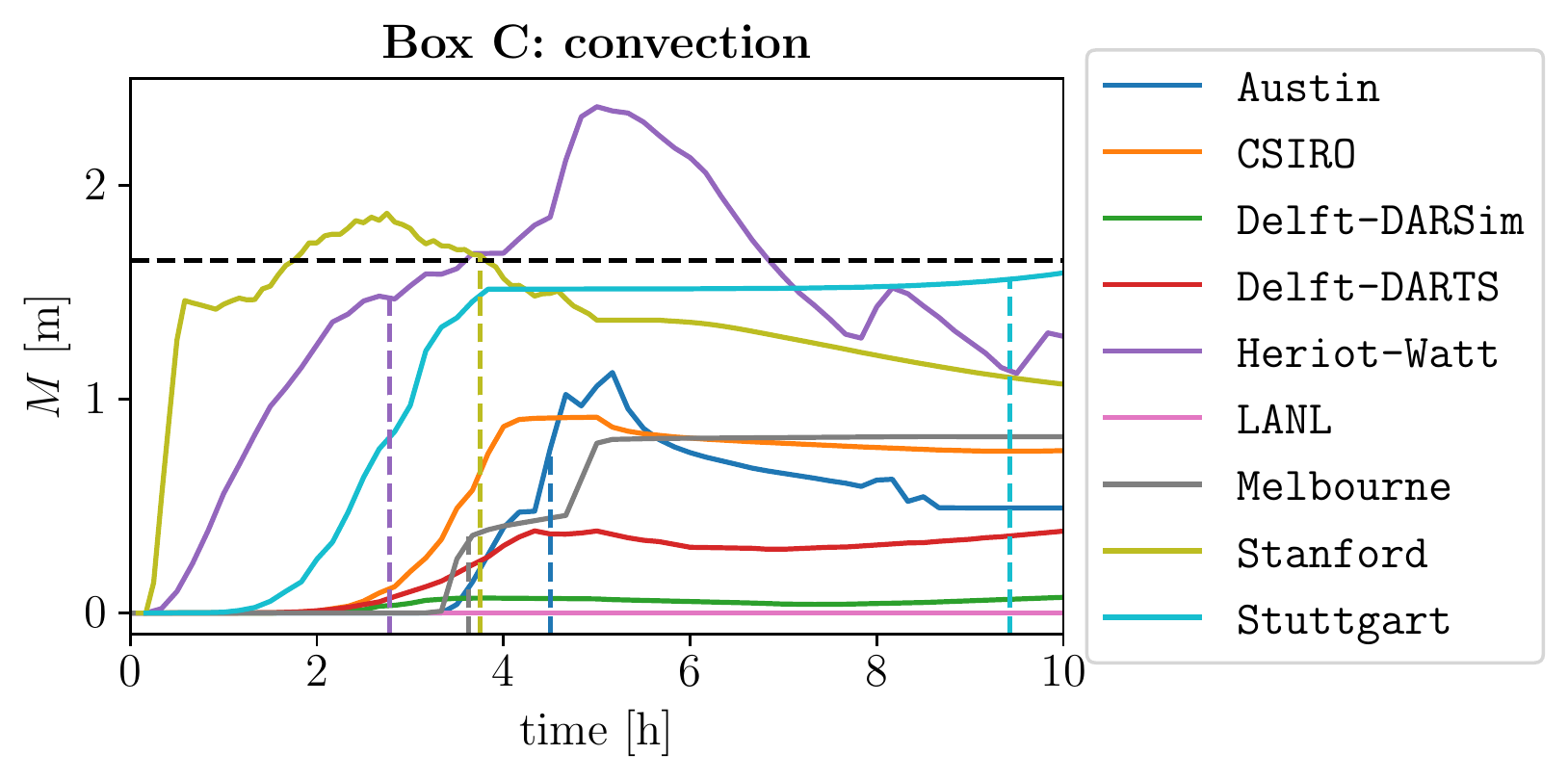}
\caption{Zoom into the first ten hours of the temporal evolution of $M(t)$. The black horizontal dashed line depicts 110\% of Box C, dashed vertical lines correspond to the reported expected mean values.}
\label{fig:time_series_boxC_zoom}
\end{figure}

As a proxy for the ability to capture the onset of convective mixing, we focus on the time for which the quantity $M(t)$ defined in \cite[Section 2.8.3]{Nordbotten:2022:FBD} first exceeds 110\% of the width of Box C, as depicted in the lower picture of Figure \ref{fig:sparse_time}. We first note that three groups don't report any value at all. Out of the remaining six, four report very similar values around \qty{4}{\hour} and narrow ranges between P10 and P90. With \csiro, one group reports much larger expected values and also variations between P10 and P90. In order to examine this in more detail, we perform a comparison with the corresponding temporal evolution of $M(t)$ as depicted in Figure \ref{fig:time_series_boxC}. With 110\% of the width of Box C being equal to \qty{1.65}{\meter}, we can observe that several results don't reach this value at all over the whole simulation period. In turn, this explains that three groups didn't report any value for the sparse data. Zooming closer into the first ten hours of simulated time as done in Figure \ref{fig:time_series_boxC_zoom} allows to put the reported time series values in explicit relation to the sparse data.
As can be identified from the vertical lines representing the reported expected mean values, the measured value for $M(t)$ is usually well below the 110\%. Therefore, it becomes obvious that several participating groups didn't rely only on the reported simulation results for the measurable considered here.

\subsubsection{Phase distributions after three days in Boxes A and B}

\begin{figure}[hbt]
\centering
\includegraphics[width=0.99\textwidth]{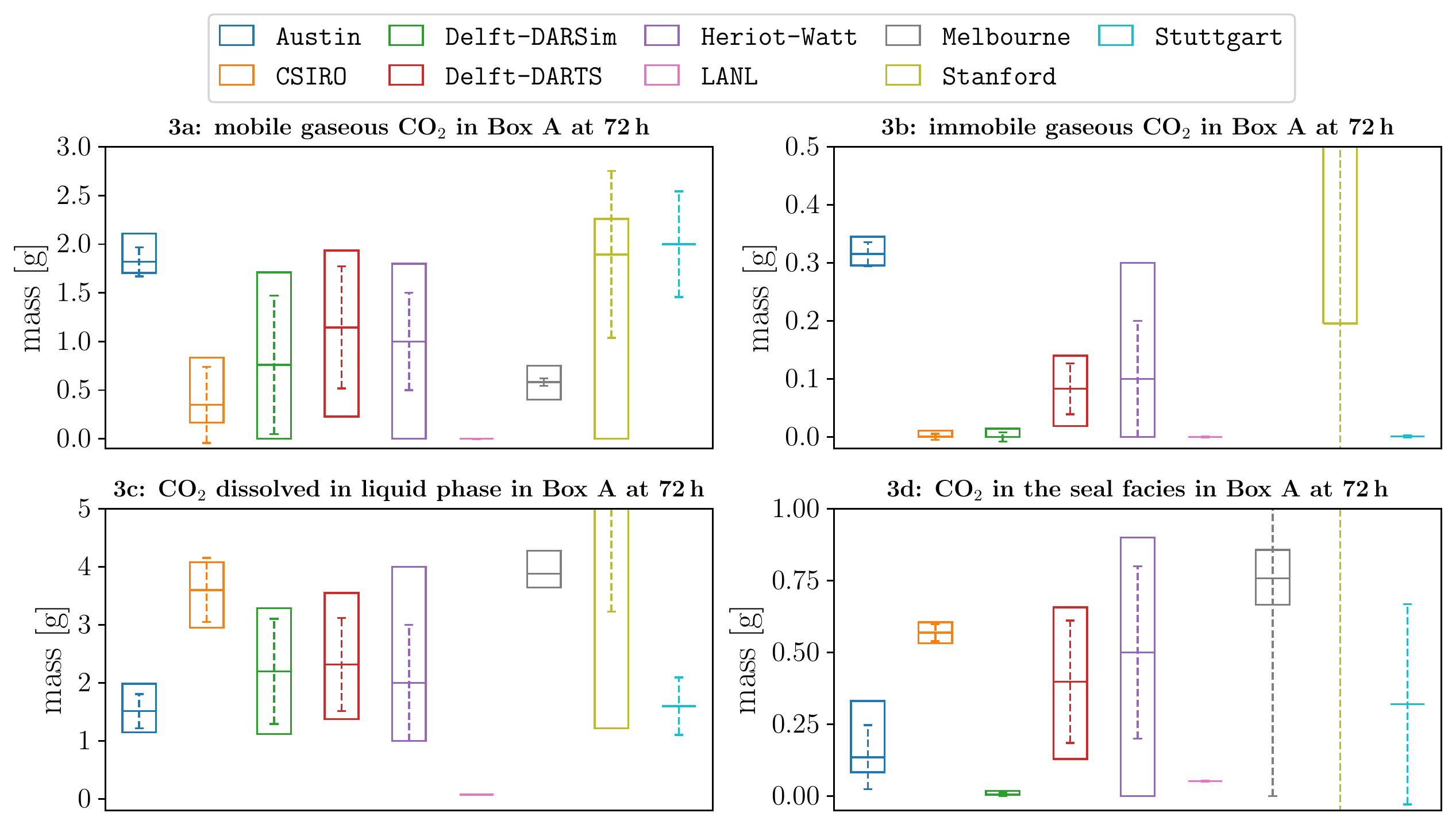}
\caption{Reported sparse data for the phase distribution in Box A at 72 hours after injection starts. Bottom, middle and top horizontal lines of the boxes indicate the reported P10, P50 and P90 values for the expected mean value, respectively. Dashed vertical lines extend from the mean values by $\pm$ the reported P50 of the expected standard deviations.}
\label{fig:sparse_boxA}
\end{figure}

We now turn to the reported sparse data for the phase distribution in Box A at 72 hours after injection starts as a proxy for the ability to accurately predict near well phase partitioning.
From the corresponding Figure \ref{fig:sparse_boxA}, it can be seen immediately that the reported ranges between the P10 and P90 values of the expected mean values are substantially larger than for the preceding measures, going along with increased expected standard deviations.
Concerning the amount of mobile gaseous \cotwo, the expected P50 of the mean value ranges between 0.5 and \qty{2}{\gram}, while for the amount of dissolved \cotwo, values range mostly between 1 and \qty{4}{\gram}.

\begin{figure}[hbt]
\centering
\includegraphics[width=0.99\textwidth]{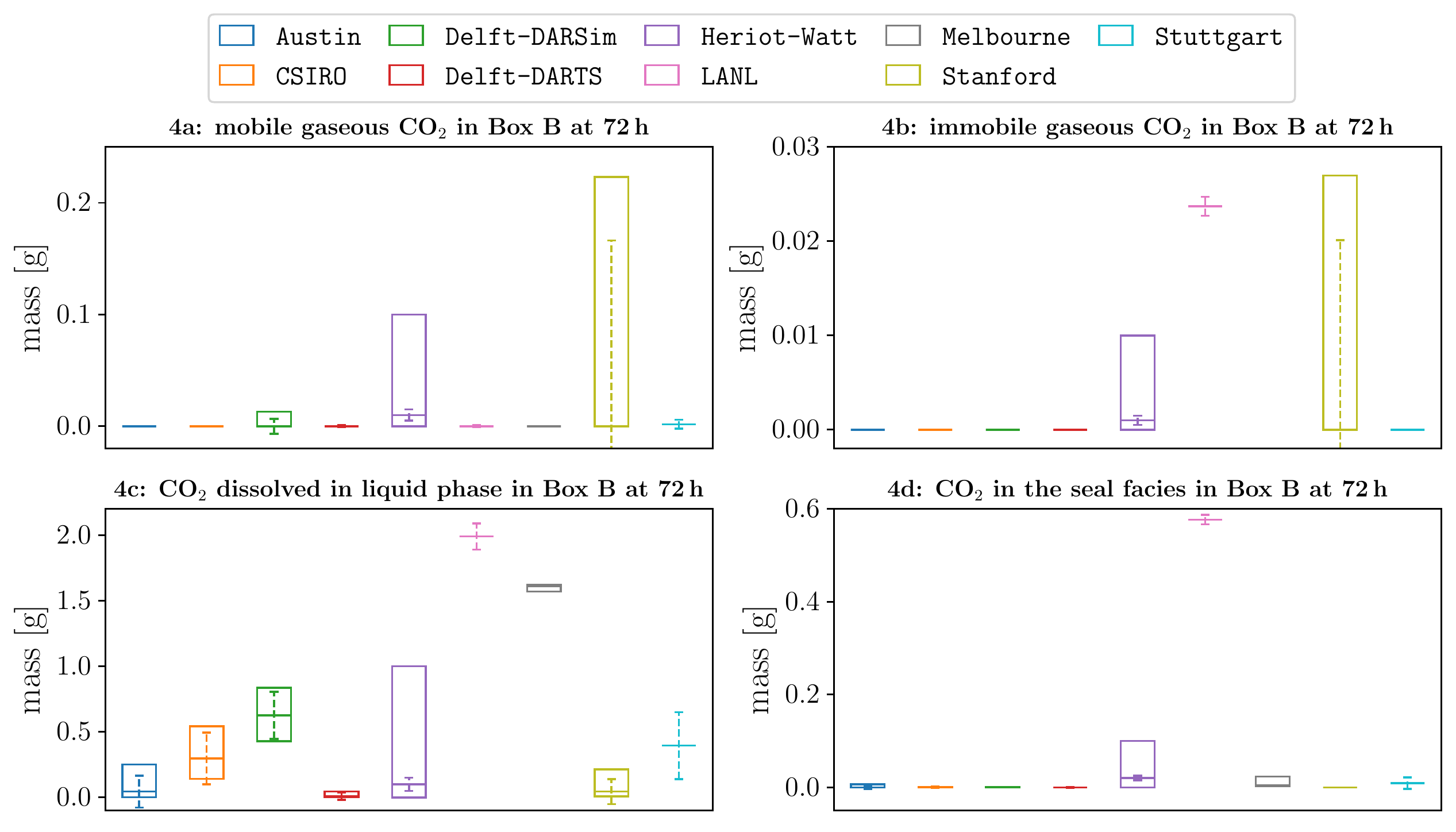}
\caption{Reported sparse data for the phase distribution in Box B at 72 hours after injection starts. Bottom, middle and top horizontal lines of the boxes indicate the reported P10, P50 and P90 values for the expected mean value, respectively. Dashed vertical lines extend from the mean values by $\pm$ the reported P50 of the expected standard deviations.}
\label{fig:sparse_boxB}
\end{figure}

The expected phase distribution in Box B at 72 hours after injection starts is depicted in Figure \ref{fig:sparse_boxB}, interpretable as a proxy for the ability to handle uncertain geological features.
It can be observed that mostly no mobile gaseous \cotwo is expected, while the associated uncertainty is considered to be quite high. In case of \stanford, the large variation comes from the fact that a simulation with immiscible fluid phases was included in the underlying uncertainty quantification as a limit case. Turning to the lower left picture, the amounts of predicted dissolved \cotwo show a strong variation over the participating groups.

\subsubsection{Total \cotwo mass in top seal facies within Box A}

\begin{figure}[hbt]
\centering
\includegraphics[width=0.99\textwidth]{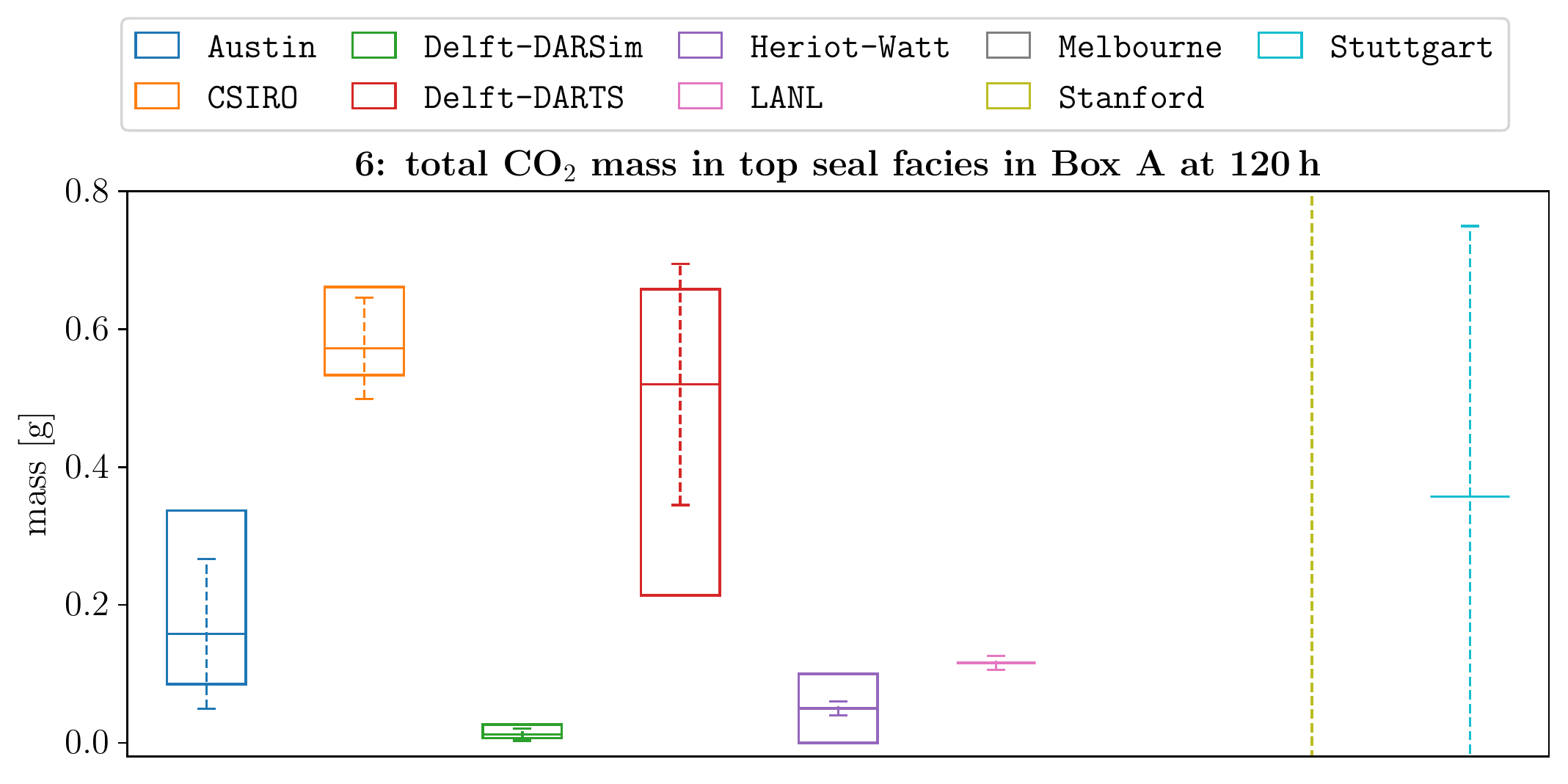}
\caption{Reported sparse data for the total mass of \cotwo in the top seal facies at final time within Box A. Bottom, middle and top horizontal lines of the boxes indicate the reported P10, P50 and P90 values for the expected mean value, respectively. Dashed vertical lines extend from the mean values by $\pm$ the reported P50 of the expected standard deviations.}
\label{fig:sparse_seal}
\end{figure}

As the last measurable, we examine the expected total mass of \cotwo in the top seal facies at final time within Box A for evaluating the ability to capture migration into low-permeable seals.
Figure \ref{fig:sparse_seal} depicts the corresponding reported results.
Also here, large variations can be observed, not only in the expected mean values, but also in the expected standard deviations. 

\section{Comparison to Experimental Data}
\label{sec:comparison}

In the following, we will compare the modeling results described in the previous section with the actually observed experimental data. The underlying experimental methodology and original dataset is presented in \cite{Ferno:2023:MCI}, while the image analysis approach is discussed in \cite{Nordbotten:2023:TIP}. We focus first on the dense data spatial maps and time series and investigate afterwards the sparse data measurables.


\subsection{Dense data spatial maps}
\label{sec:comp_spatial}

\begin{figure}[hbt]
\centering
\includegraphics[width=0.32\textwidth]{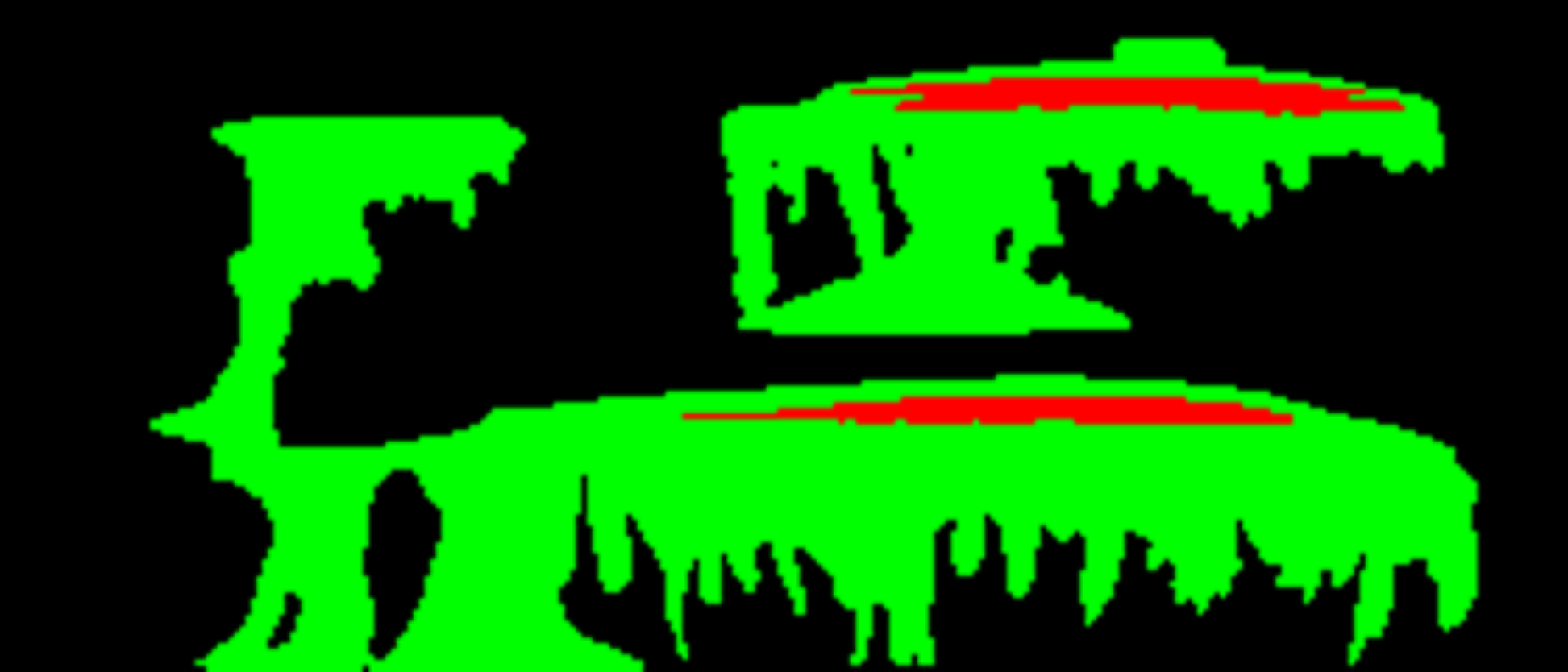}
\includegraphics[width=0.32\textwidth]{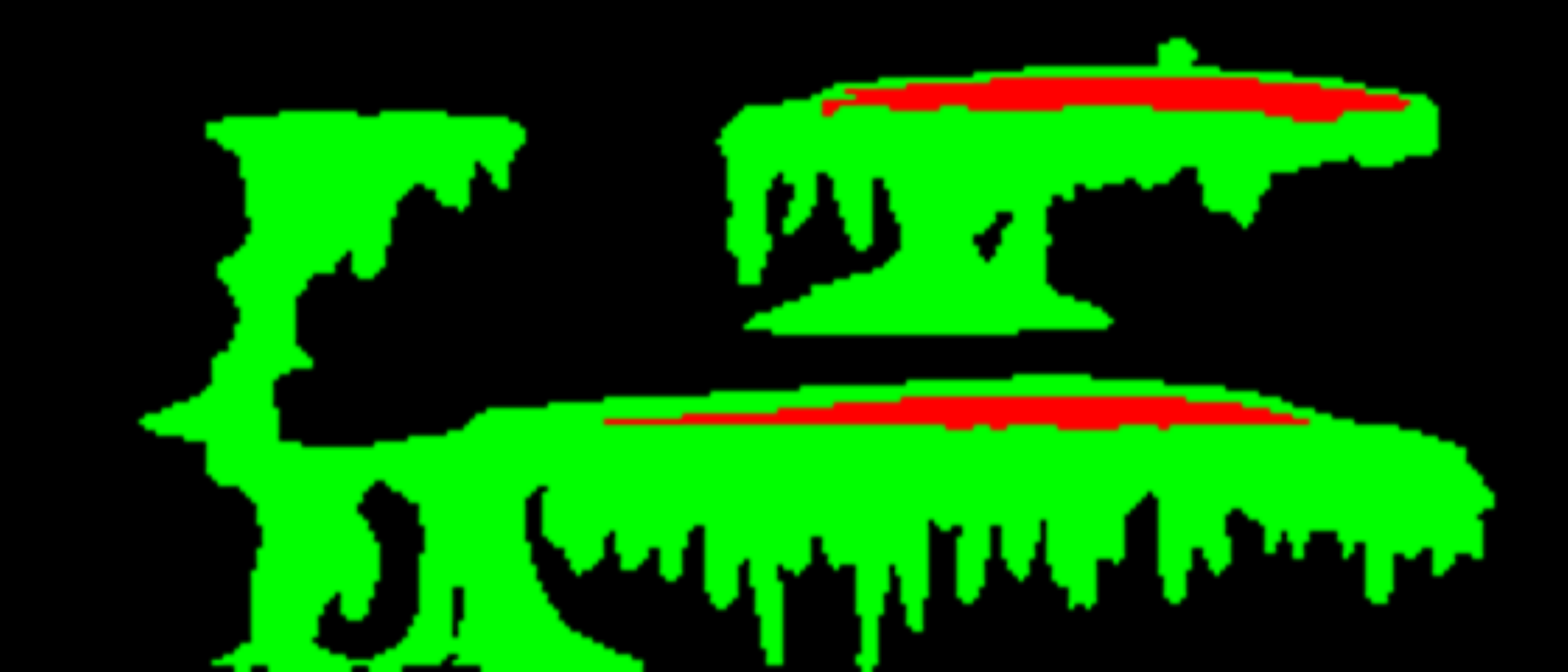}\\
\includegraphics[width=0.32\textwidth]{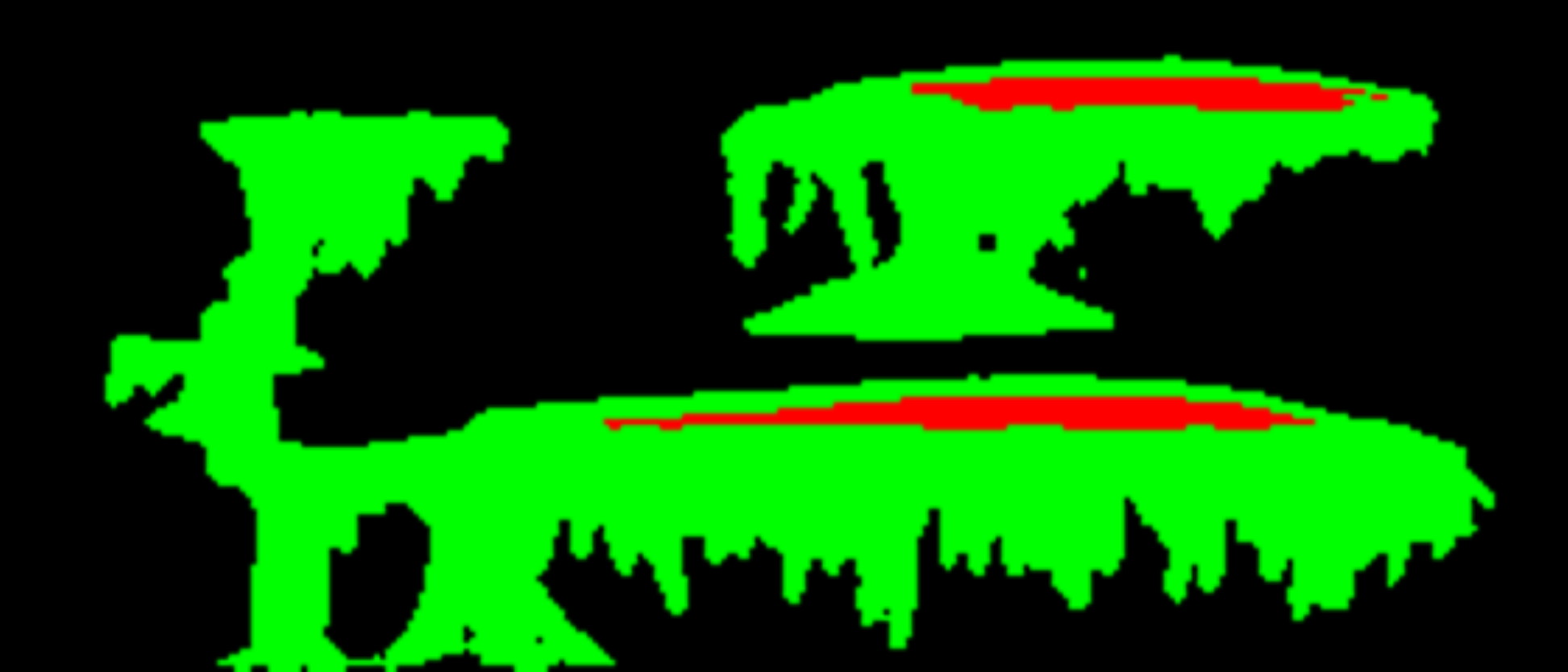}
\includegraphics[width=0.32\textwidth]{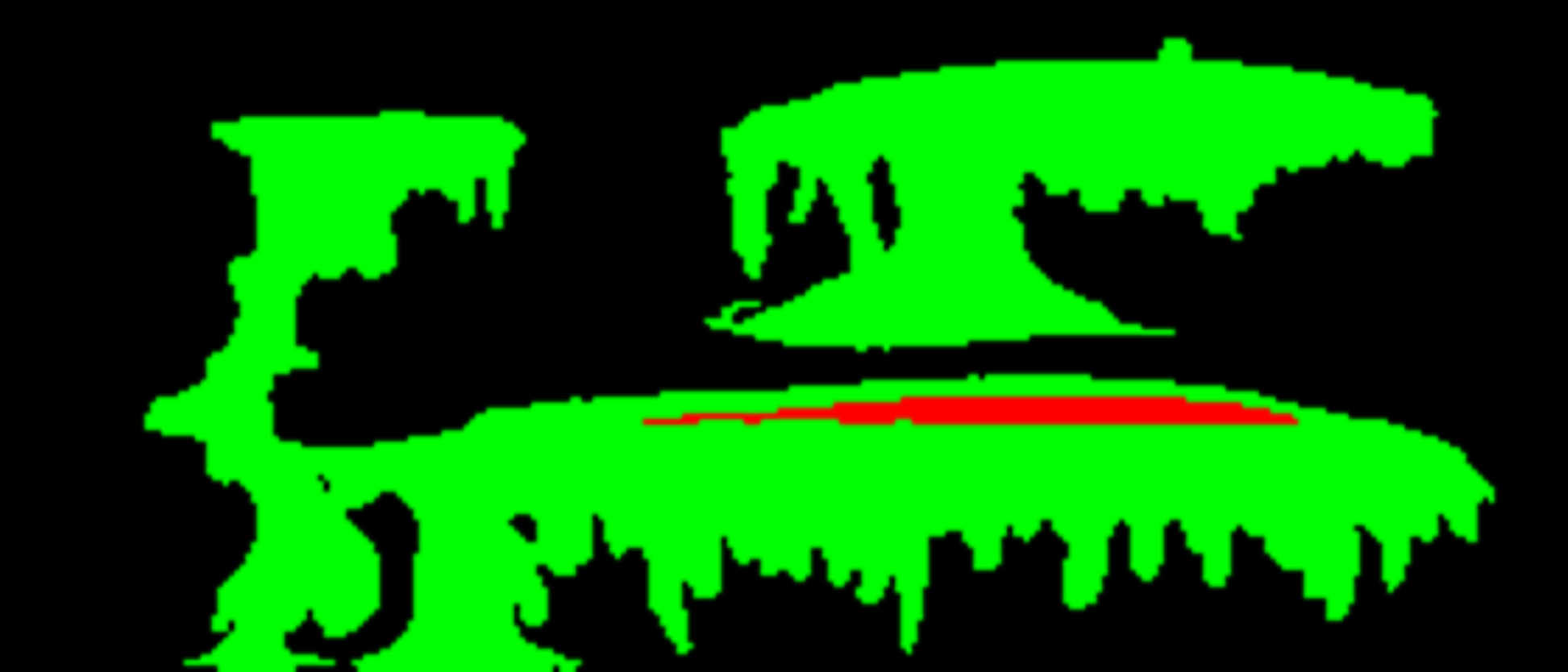}
\includegraphics[width=0.32\textwidth]{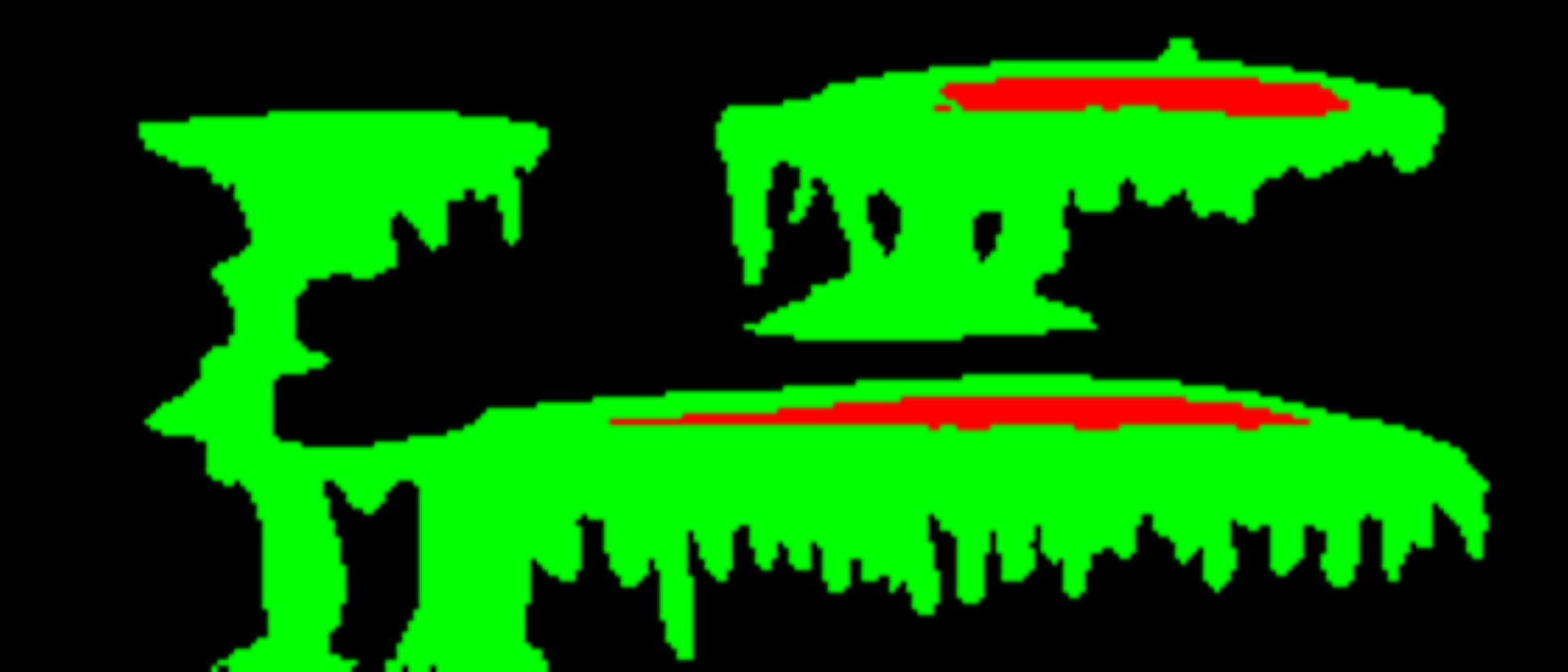}
\caption{Segmentation data after \qty{24}{\hour} for five experimental runs. Black, green and red indicate pure water, water with dissolved \cotwo and gas, respectively.}
\label{fig:spatial_maps_exp}
\end{figure}

In the following, we compare daily spatial maps given in form of segmentation data. For the experiments, this data has been generated by analyzing corresponding images using the newly developed toolbox DarSIA~\cite{Nordbotten:2023:TIP}. In Figure \ref{fig:spatial_maps_exp}, the snapshots at \qty{24}{\hour} are shown for five experimental runs.
Visually, there is a very good agreement over all five runs and differences can only be detected in the details. One slight exception is given by the fourth run, where no gas appears to be present in the upper right part of the domain. However, this is attributable to numerical effects in the image analysis procedure, rather than a different physical truth. We will perform a quantitative analysis further below.

\begin{figure}[hbt]
\centering
\includegraphics[width=0.99\textwidth]{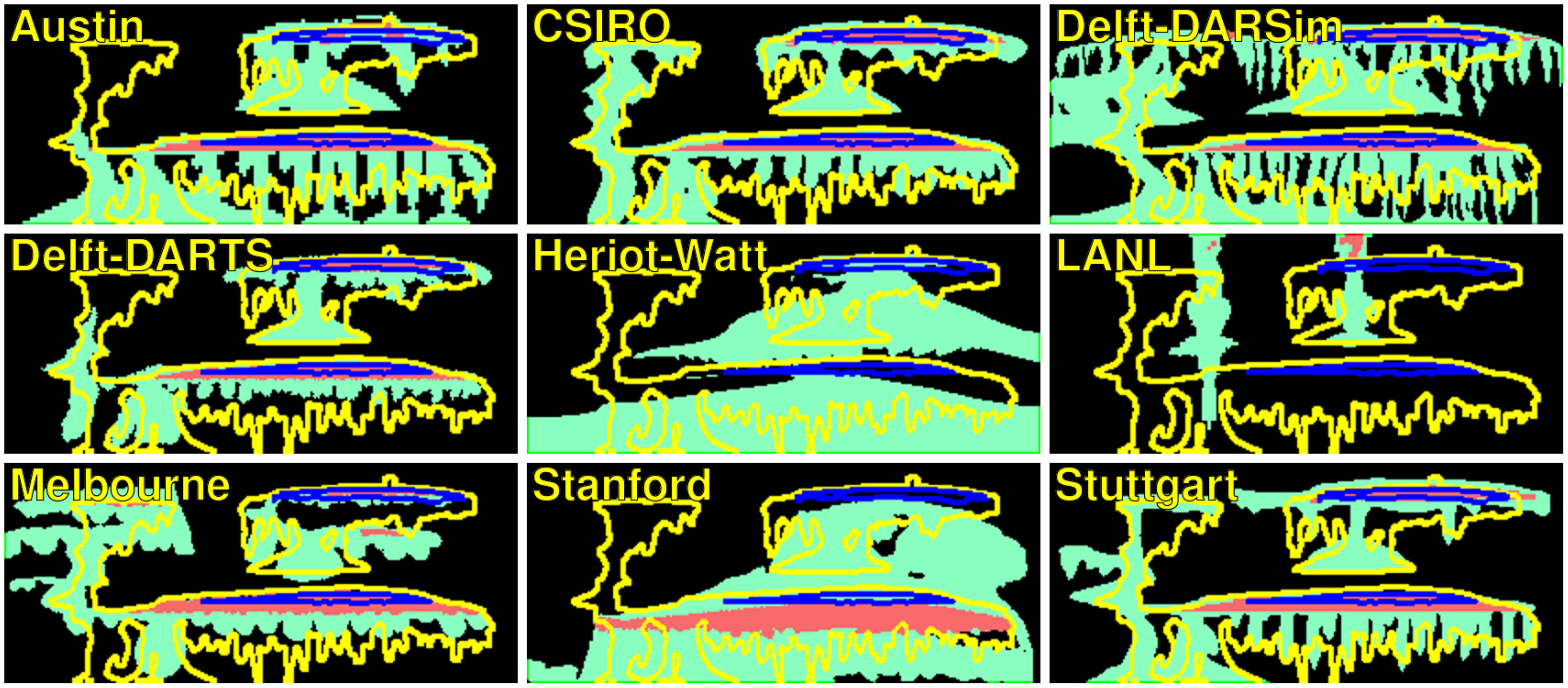}
\caption{Comparison of segmentation data after \qty{24}{\hour}. Each modeling result is overlaid by the contour lines of experimental run 2. The forecasts are colored by black, pale green and pale red, indicating pure water, water with dissolved \cotwo and gas, respectively. Concerning the experimental data, yellow contour lines indicate the region of water with dissolved \cotwo, while blue lines illustrate the gas plume.}
\label{fig:compare_seg}
\end{figure}

Before that, a visual comparison with the modeling results is carried out. For this, the concentration and saturation maps at \qty{24}{\hour} provided by the participants are converted into segmentation data. To allow for a more direct comparison, the modeling results are overlaid by the contour lines corresponding to the experimental data. The result is shown in Figure \ref{fig:compare_seg}.
It  can be seen that the locations of the two gas plumes are reasonably well captured by several models, namely, \austin, \csiro, \darsim, \darts, \melbourne and \stuttgart, while their sizes are overestimated in general. As already suggested by the strong variability of the concentration distributions discussed in Section \ref{sec:results_spatial_maps}, considerably less agreement can be observed concerning the region covered by water with dissolved \cotwo. This becomes particularly apparent for Box B in the upper left part of the domain, where only the \csiro modeling result matches the basic shape and extension in a visually satisfactory way.

\begin{figure}[hbt]
\centering
\includegraphics[width=0.99\textwidth]{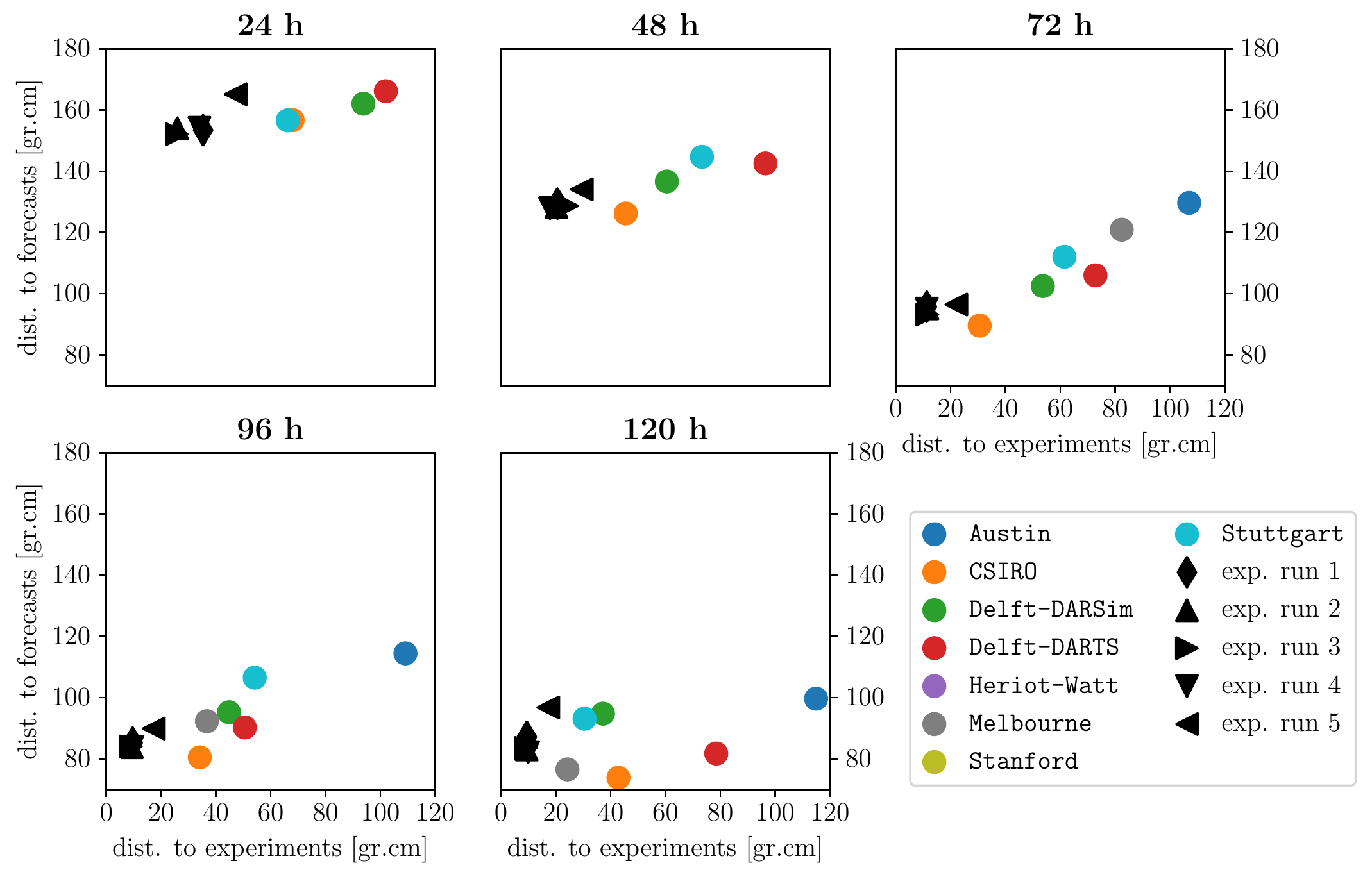}
\caption{Wasserstein distances of the segmentation maps to experiments and forecasts. Zoom into the ranges from 0 to 120 gr.cm for the mean distance to the experimental results and from 70 to 180 gr.cm for the mean distance to the modeling forecasts. Some groups with outlying results are therefore not visible in all plots, while \hw and \stanford are consistently outside the range of the plots (confer distances in Figure \ref{fig:pcolor}).}
\label{fig:means_segmented_zoom}
\end{figure}

To develop a more quantitative understanding, a similar analysis as in Section~\ref{sec:wasserstein_models} can be performed in terms of the Wasserstein metric. This involves calculating distances for all pairs consisting of two participating groups, two experimental runs, or one participant and one run. Similar to above, the mean distances to the other modeling results and now also to the experimental data can be calculated, yielding two values for each segmentation map. Figure \ref{fig:means_segmented_zoom} plots these values for all segmentation maps at the selected time steps.
We can observe that the experimental data sets are within 50 gr.cm of each other, confirming that the experimental repeatability is strong, and that there is only minor impact of the different experimental conditions (primarily attributed to atmospheric pressure, some chemical alterations within the experimental rig, and a very minor amounts  settling of sand throughout the experimental period). About half of the modeling results are within about 100 gr.cm of the experimental data for all reporting times, which we consider a relatively good match. At the final time, the closest simulation results are as little as 50 gr.cm away from the experimental mean, which is within twice the experimental variability at that time. This also aligns with the visual impressions for the segmented images shared above. With increasing time, the distances to both the experiments and the forecasts are decreasing for most modeling results; the same holds for the distances of the experimental data sets to the forecasts. This can be explained by the increasing spread of \cotwo-rich water over the domain and a corresponding equilibration of \cotwo mass.

\subsection{Dense data time series}

In the following, we compare selected dense data time series as reported by the participating groups with corresponding experimental data. As described in \cite{Ferno:2023:MCI,Nordbotten:2023:TIP}, the derivation of saturation and concentration values from the experimental photographs is a very challenging endeavor based on several assumptions. The correspondingly calculated mass values are subject to significant uncertainties. Therefore, the degree of physical truth behind the comparisons has to be taken with great care.

\begin{figure}[hbt]
\centering
\includegraphics[width=0.99\textwidth]{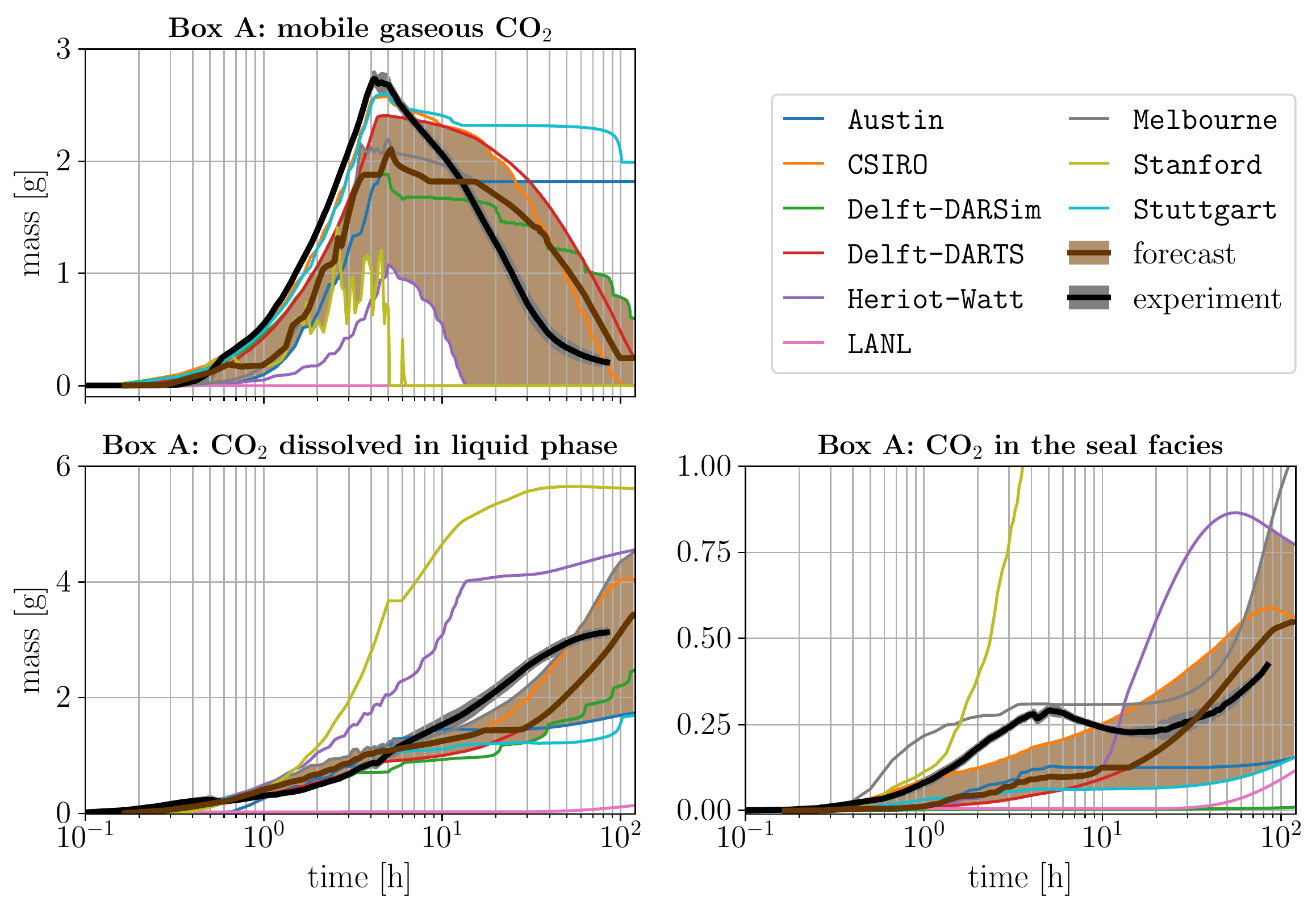}
\caption{Comparison between modeling forecasts and experimental observations for the temporal evolution of the \cotwo phase distribution in Box A. A brown line depicts the median of the reported modeling results, while the associated pale brown region illustrates the area between the corresponding first and the third quartile. A black line shows the mean of the experimental data, while the associated grey region depicts the corresponding variation by means of the standard deviation.}
\label{fig:compare_time_boxA}
\end{figure}

Figure \ref{fig:compare_time_boxA} shows the comparison for the temporal evolution of the phase distribution in Box A.
For being able to observe more details in the beginning of the investigated time frame, the x-axes in the pictures use a logarithmic scaling.
Concerning mobile gaseous \cotwo, the basic shape of the experimental mean is quite similar to the median of the modeling results. Nevertheless, the peak value for the forecast is considerably lower than the experimental one. The spread of the modeling results during the advection-driven stage of increasing values is substantially less than during the dissolution-driven stage of decreasing values afterwards. This results in a much longer period where the value stays rather constant. While in general the stages of increasing and decreasing values are lagging behind the experimental results, the results from \csiro and \stuttgart match the first stage very well.

Focusing on the temporal behavior of the dissolved \cotwo mass, it can be seen that most of the modeling results agree well with the experimental data in the beginning. The spread in the forecasts starts to increase after the injection stops and the very different dissolution behaviors discussed earlier become dominant. While most modeling results underestimate the amount of dissolved \cotwo during the majority of the simulated time, the values tend to increase longer than the corresponding experimental data which saturates earlier. Investigating the third picture, the evolution of the \cotwo mass in the seal varies strongly over the participating groups and differs substantially from the experimental data. A possible reason for the non-monotonic behavior of the experimental mean is discussed in \cite{Ferno:2023:MCI}.

\begin{figure}[hbt]
\centering
\includegraphics[width=0.99\textwidth]{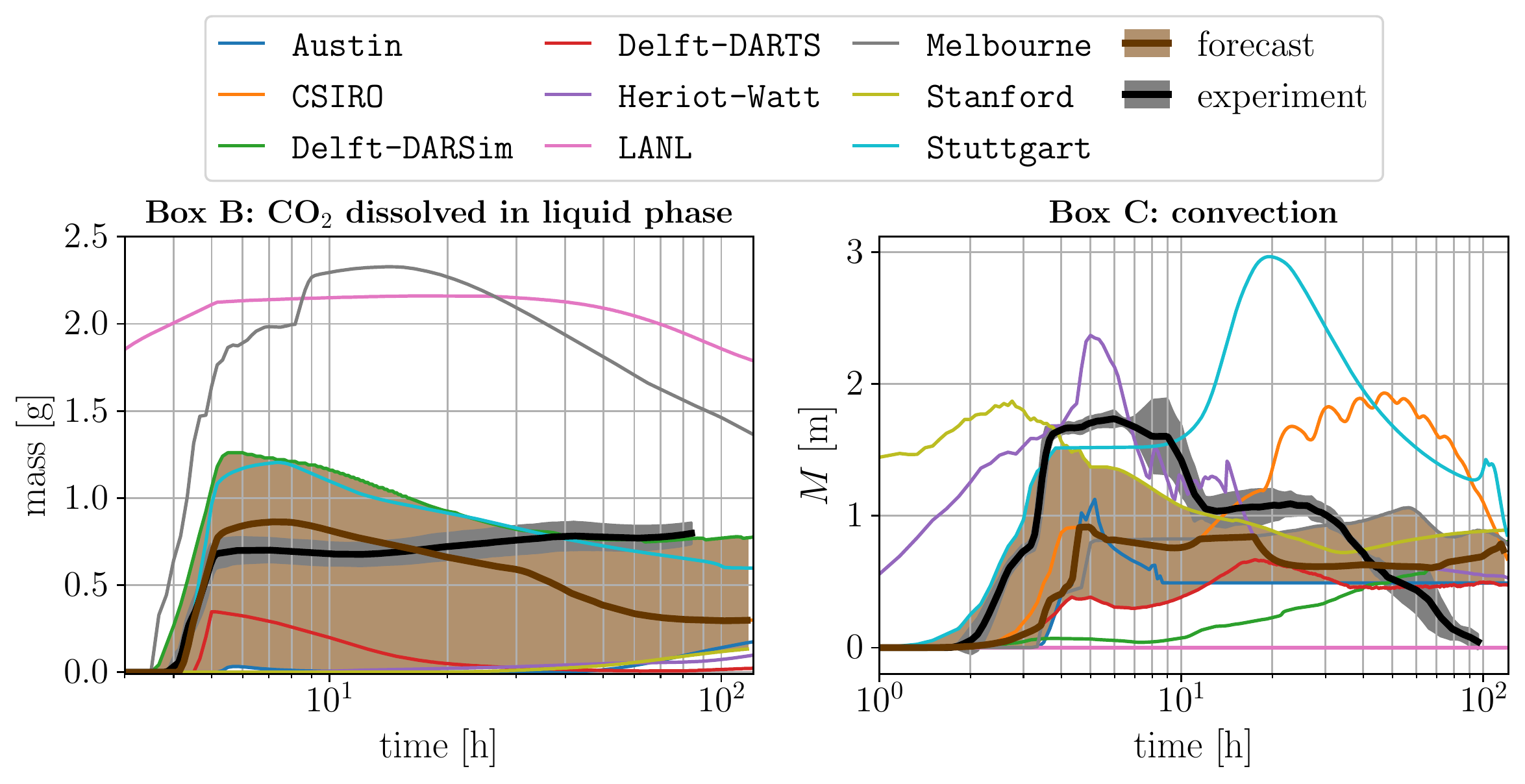}
\caption{Comparison between modeling forecasts and experimental observations for the temporal evolution of the dissolved \cotwo mass in Box B (left) and the integral quantity $M(t)$ (right). A brown line depicts the median of the reported modeling results, while the associated pale brown region illustrates the area between the corresponding first and the third quartile. A black line shows the mean of the experimental data, while the associated grey region depicts the corresponding variation by means of the standard deviation.}
\label{fig:compare_time_boxBC}
\end{figure}

Experimental data has been provided for two other time series and the corresponding comparisons are illustrated in Figure \ref{fig:compare_time_boxBC}.
Turning first to the amount of dissolved \cotwo in Box B, the large variations in the modeling results are also apparent by the depicted large spread. Like for Box A, the advection-driven increase in the beginning is captured well by two participating groups. Also here, the differences become more pronounced after injection stops. The amount of \cotwo increases further in the experimental data over time due to \cotwo-rich water entering Box B from the right. This effect is not captured by most of the models.

We investigate finally the temporal evolution of the convection measure $M(t)$ in the right picture of Figure \ref{fig:compare_time_boxBC}. However, the differences of the modeling results to the experimental data are too strong to draw any meaningful conclusion here. It is likely that this has to do with the fact that the numerical evaluation of the integral value is not straightforward, strongly discretization-dependent and has been left entirely to the participants.

\subsection{Sparse data}

The collection of the sparse data results has been accompanied by questionnaires for monitoring the confidence of each participant in their own prediction as well as in the ones of the respective other working groups. Since the description and analysis of this process and its results would be beyond the scope of this work, a separate paper is devoted to this~\cite{Nordbotten:2023:OBP}. In the following comparison with the experimental data, we therefore limit ourselves to a rather brief presentation of a few agglomerated measures.

In order to condense the responses by the individual participating groups presented in Section \ref{sec:results_sparse_data}, we only consider the reported P50 values for the expected means and standard deviations. The means will be plotted as individual data points, together with their median and the median of the expected standard deviations. Concerning the experimental data, the results from the individual runs are plotted, together with their mean and standard deviation.

\begin{figure}[hbt]
\centering
\includegraphics[width=0.8\textwidth]{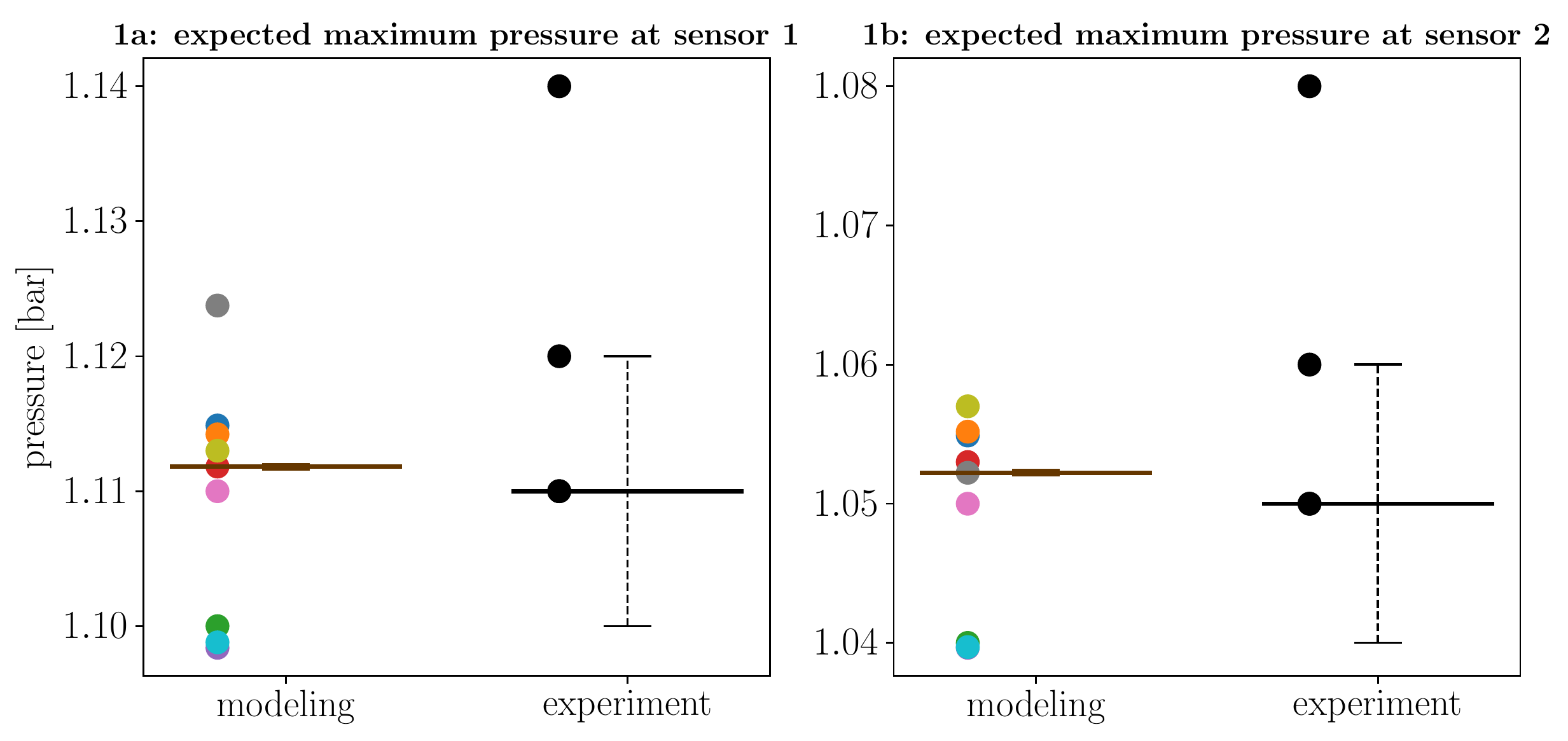}
\caption{Comparison of the sparse data reported by the participating groups with the experimental data for measurable 1. Concerning the modeling results, colored circles correspond to the individual expected means, while the horizontal brown line depicts their median. A dashed vertical brown line extends from this value by $\pm$ the median of all reported P50 values for the standard deviation. Some outlier values might be outside the plotting range. Regarding the experimental data, black circles depict the results of the individual runs, while the horizontal black line indicates their mean. A dashed vertical black line extends from the mean by $\pm$ the standard deviation.}
\label{fig:compare_sparse_pressure}
\end{figure}

In Figure \ref{fig:compare_sparse_pressure}, we consider first with measurable 1 the expected and observed maximum pressures in the two sensors.
Like predicted by most of the participating groups, the injection of \cotwo had almost no impact on the pressure observed in the two sensors. The reported measured experimental values correspond to the maximum atmospheric pressure during a respective experimental run plus the hydrostatic contribution by the corresponding overlying water column. The individually reported expected means are within \qty{10}{\milli\bar} of the experimental mean and the median of the expected means shows a very good agreement with the experimental mean. Nevertheless, as already noticed in Section \ref{sec:maxpress}, the participants expected almost no variation in the experimental results. Due to the natural fluctuations in atmospheric pressure, the observed variations turn out to be significantly larger than the expected ones.

\begin{figure}[hbt]
\centering
\includegraphics[width=0.8\textwidth]{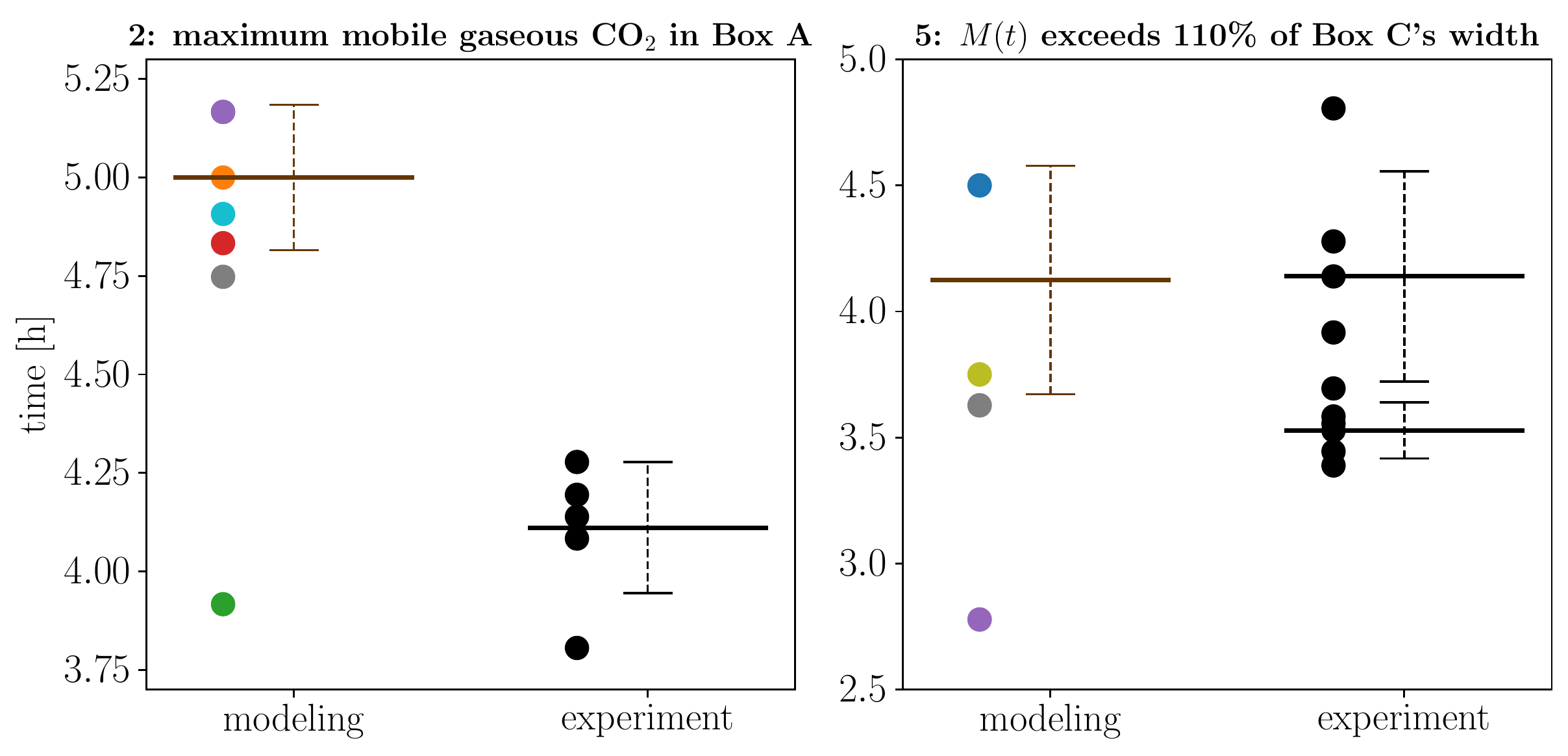}
\caption{Comparison of the sparse data reported by the participating groups with the experimental data for measurables 2 (left) and 5 (right). See Figure \ref{fig:compare_sparse_pressure} for more details on the plotted quantities.}
\label{fig:compare_sparse_time}
\end{figure}

Figure \ref{fig:compare_sparse_time} illustrates the comparison for the measurables 2 and 5, namely, the time of maximum mobile gas phase in Box A and the time when $M(t)$ exceeds 110\% of Box C's width, respectively.
Concerning the former, it can be observed that the experimental mean is overestimated by most participating groups and that the reported and observed ranges are rather disjoint. For the latter, the situation is different as two sets of experimental data are provided which differ in the underlying image analysis parameters and constitute upper and lower bounds for the target quantity. Here, the median of the expected means lies close to the corresponding upper experimental mean. 

\begin{figure}[hbt]
\centering
\includegraphics[width=0.99\textwidth]{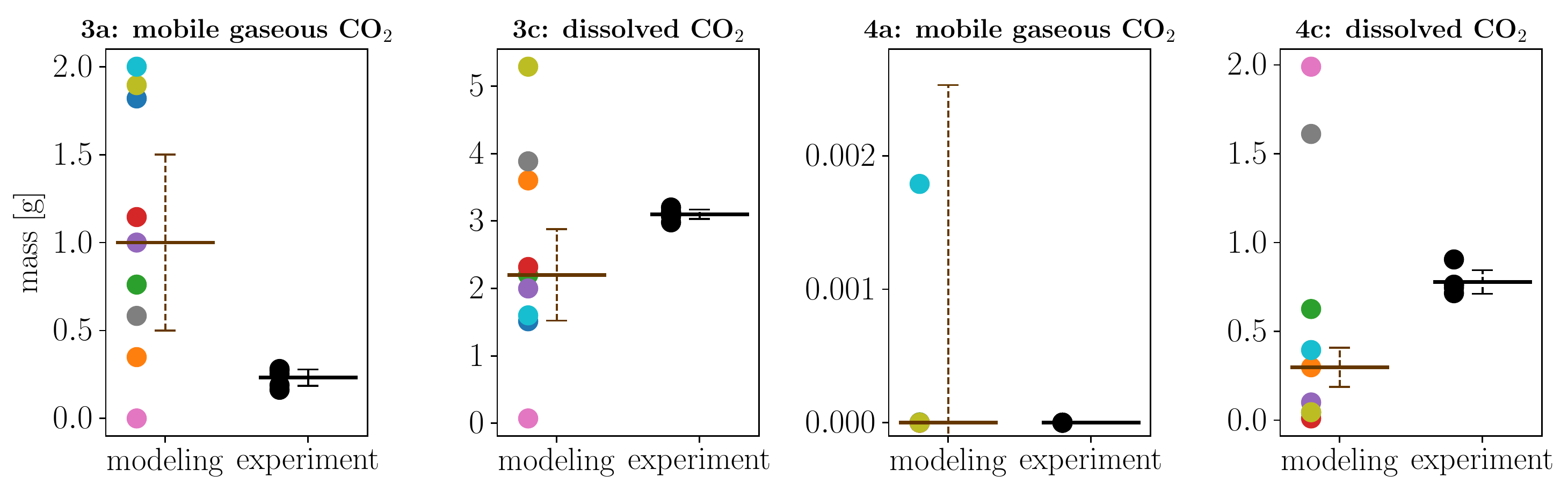}
\caption{Comparison of the sparse data reported by the participating groups with the experimental data for measurables 3a, 3c, 4a and 4c (left to right). See Figure \ref{fig:compare_sparse_pressure} for more details on the plotted quantities.}
\label{fig:compare_sparse_mass}
\end{figure}

Next, we perform a comparison for the sparse data measurables 3a, 3c, 4a and 4c, regarding the phase distribution of \cotwo after \qty{72}{\hour} in Box A and B, respectively.
Figure \ref{fig:compare_sparse_mass} depicts the corresponding quantities in terms of \cotwo mass in either gaseous or liquid phase.
Starting with 3a, it can be observed that the mean value of mobile gaseous \cotwo in Box A is overestimated by most participating groups and only some groups report values within the observed experimental range. This is consistent with the visual impressions discussed in Section \ref{sec:comp_spatial}. Regarding 3c, the mean value of \cotwo dissolved in water in Box A is rather underestimated by the modelers. Moving to Box B, all experimental runs suggest that no gaseous \cotwo is left after \qty{72}{\hour}. This has also been expected by most participants, while they nevertheless presumed a slight standard deviation on average. While the reported numbers for the expected mean of dissolved \cotwo are rather widespread, the median value is remarkably close to the observed experimental mean.

\begin{figure}[hbt]
\centering
\includegraphics[width=0.4\textwidth]{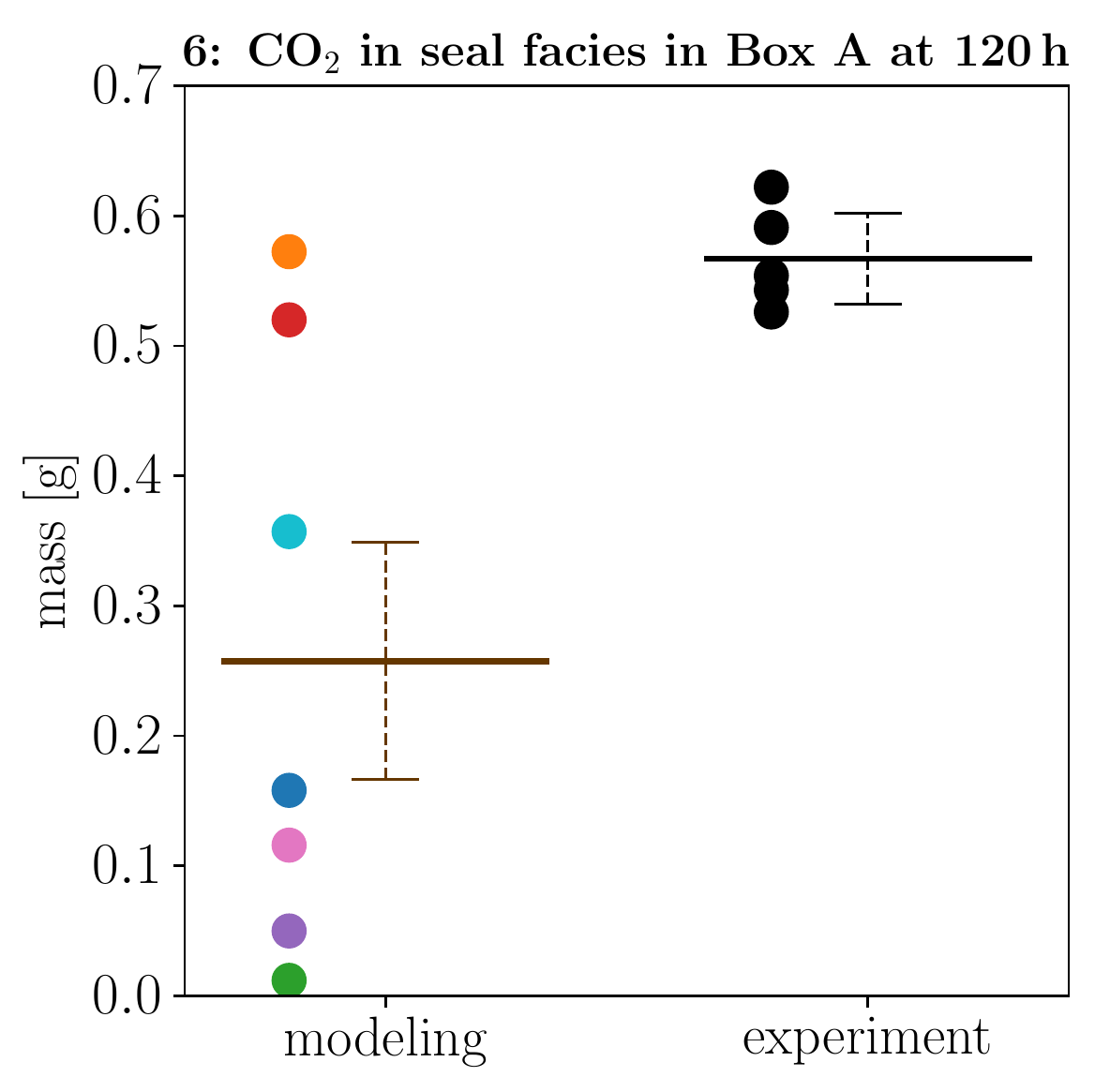}
\caption{Comparison of the sparse data reported by the participating groups with the experimental data for measurable 6. See Figure \ref{fig:compare_sparse_pressure} for more details on the plotted quantities.}
\label{fig:compare_sparse_seal}
\end{figure}

With the final measurable 6, we examine the total \cotwo mass in top seal facies within Box A at final simulation time, as illustrated in Figure \ref{fig:compare_sparse_seal}.
The median of the expected means is at around 50\% of the observed experimental mean. Correspondingly, most participating groups underestimate the amount of \cotwo in the top seal facies. Nevertheless, two groups are very close to the experimental results.

\section{Conclusion and Outlook}
\label{sec:conclusion}

In the following, we will draw several conclusions from this validation benchmark study and present challenges and opportunities for further work.

First, we can state with strong confidence that Darcy-scale balance equations together with standard constitutive relationships for the capillary pressure and relative permeability describe adequately the relevant observed physical processes on the considered spatial and temporal scale. This is revealed clearly from the comparison of the modeled saturation and concentration distributions with the corresponding experimental segmentation maps. In particular, stratigraphic and residual trapping mechanisms are captured well by most participating groups. Moreover, the process of convective mixing due to density differences is considered adequately in a qualitative manner.

Quantitatively, large variations in the modeling results can be observed particularly for the dissolution behavior and the resulting fingering. This can be attributed to different modeling choices for the solubility limit of \cotwo in water as well as for constitutive relations such as capillary pressure - saturation relationships, equations of state for determining phase compositions or phase density calculations. It can also be observed that differences in grid resolution clearly influence the convective mixing behavior. Nevertheless, several participating groups are in close proximity to the experimental results, as quantified by the Wasserstein metric. The corresponding distances decrease with increasing time as more \cotwo is dissolved and its mass equilibrates over the domain. 

The study included reporting of pre-defined ``sparse data'', which were quantities that we can consider as proxies for various aspects of storage capacity and storage security. These quantities were reported with both a most likely exceedance value (P50), as well as P10-P90 intervals. While the P50 values mostly reproduce the reported dense data, the P10-P90 values add an additional dimension to the results. Notably, for the majority of requested quantities, the reported P10-P90 quantities do not overlap between the groups. Logically speaking, if two P10-P90 intervals do not overlap, then one group believes that there is at most a 10\% chance that the other group will find the experimental results to be within their reported interval (and conversely). This implies that despite the significant group interaction through the study, the groups did not take the quantitative response of other groups into serious consideration, and placed high or full confidence in their own results. This observation is complemented by the fact that the interaction helped very well almost all groups to establish a common understanding regarding the expected qualitative behavior such as the effect of capillary barriers.

A particular critical physical process that is evidenced in this study (both in sparse and dense data) is the role of convective mixing in accelerating dissolution of gaseous CO$_2$. This is quantified both through the actual phase compositions in Box A and B, as well as in the metric $M$, which is a proxy for the time of fully developed fingers (for a detailed discussion of various onset times in numerical simulation of density driven fingers, see~\cite{Elenius:2012:TSN}.  The onset and evolution of convective fingers is particularly challenging for this system, since the low-order numerical methods used in this study (suitable to capture heterogeneity and stable discretization of multi-phase flow) tend to be too diffusive in their representation of the gas-water interface. The result is significantly over-estimating mass transfer from the gas to the water phase, necessitating a fine grid in the vertical direction. Moreover, the characteristic wave-length of density driven fingers for this system is on the order of 5 cm (as seen experimentally), requiring further necessitating a sub-centimeter grid resolution horizontally.  Seen together, this may be the cause for large variability in the reported structure and importance of density-driven fingering among the participants, and motivates further study on how to reliably and accurately capture this process within reservoir simulation tools.


While this study is at the laboratory scale, the fundamental physical processes of multi-phase, multi-component flows in heterogeneous porous media are the same as at reservoir conditions. As such, we argue that the findings and observations in this study are indicative of field-scale simulation (for a detailed scaling analysis, see Kovscek et al, this volume). That said, actual field-scale simulation will deviate from this study in several important aspects, of which we highlight:

\begin{itemize}
    \item Heterogeneity. This study was conducted with homogeneous facies (to the extent possible in laboratory conditions), emphasizing larger scale structural heterogenities. On the field scale, it is expected that there will be significant subscale heterogeneity also within each geological stucture. 
    \item Quality of geological characterization. This study was conducted in a quasi-2D geometry, which was fairly well characterized (high-resolution photography as well as thickness measurements at the beginning of the experiment). At the field scale, the geological characterization is based on seismic surveys, which are not able to provide the same level of accuracy. 
    \item Dimensionality. Reality is 3D, which will impact simulation time, and thus indirectly the level of grid refinement that can be sought.
    \item Convective mixing. In field-scale simulations, the spatial and temporal resolutions required for capturing correctly convective mixing are not practically feasible.
    \item Pressure and temperature conditions. At laboratory conditions CO$_2$ exists in a gas phase, while at field scale typically reservoirs with pressure and temperature compatible with supercritical CO$_2$ is sought. This has a minor impact on viscosity, but leads to a denser and less compressible CO$_2$ phase. 
\end{itemize}

What actually is very different from reservior conditions at depth is the importance of pressure measurements. In the experiment, pressure signals are rather uninformative and might introduce differences in permeability interpretation, whereas they are valuable in a reservoir context. Another major consideration is that the subsurface is much harder to characterize than the experimental rig, and so the uncertainties in predictions are going to dominated by uncertainties in geological characterisation. This code comparison illustrates the range of predictions that are possible in a relatively well-characterised system. 

From a reservoir simulation perspective, all participants reported that they struggled to achieve acceptable run times, and were forced to use relatively coarse grids for this study. We speculate that this is due to the low density of the gas phase, which has the consequence that when CO$_2$ dissolves into water, the resulting mixture has significantly lower volume than before mixing. This study thus provides impetus for further development of efficient non-linear solvers for soluble gas-water systems.   


\backmatter


\bmhead{Acknowledgments}
B.~Flemisch thanks the German Research Foundation (DFG) for supporting this work by funding SFB 1313, Project Number 327154368.
S.~Geiger acknowledges partial funding from Energi Simulation. 
H.~Hajibeygi was sponsored by the Dutch National Science Foundation (NWO) under Vidi Talent Program Project ``ADMIRE'' (Project Number 17509).

\bibliography{fluidflower_ibs}

\section*{Statements and Declarations}

\subsection*{Funding}
S.~Geiger acknowledges partial funding from Energi Simulation. 
H.~Hajibeygi was sponsored by the Dutch National Science Foundation (NWO) under Vidi Talent Program Project ``ADMIRE'' (Project Number 17509).

\subsection*{Competing interests}
The authors have no relevant financial or non-financial interests to disclose.

\subsection*{Author Contributions}
B.~Flemisch, J.~Nordbotten, M.~Fernø and R.~Juanes conceptualized, designed and implemented the benchmark study. All other authors constitute the participating groups and correspondingly set up, executed and evaluated the simulations and provided the requested results, together with descriptions of the underlying models. B.~Flemisch and J.~Nordbotten wrote the initial draft of the manuscript, all other authors were involved in the internal review and editing process. B.~Flemisch wrote and executed the scripts for generating all figures in the manuscript. All authors read and approved the final manuscript.

\subsection*{Data Availability}
All data which has been used for generating the figures in this paper is collected in respective repositories of the GitHub ``FluidFlower'' organization, which is accessible at \href{https://github.com/fluidflower}{github.com/fluidflower}. In particular, the results provided by the participating groups are collected in repositories \url{github.com/fluidflower/}\emph{groupname}\url{.git}, where \emph{groupname} is out of \url{austin}, \url{csiro}, \url{delft}, \url{heriot-watt}, \url{lanl}, \url{melbourne}, \url{stanford} and \url{stuttgart}. The experimental data used for comparison with the modeling results is assembled in  \href{https://github.com/fluidflower/experiment}{github.com/fluidflower/experiment}. The scripts for the generation of all figures are contained in \href{https://github.com/fluidflower/general}{github.com/fluidflower/general}.

\begin{appendices}

\section{Wasserstein distances}\label{sec:distances}
The following tables list the Wasserstein distances between the spatial maps as provided by the participating groups for each requested timestep. For the calculation, the Python library POT~\cite{Flamary:2021:POT} has been used. The full data including distances between results from different timesteps is provided in the FluidFlower general GitHub repository. For obtaining the numbers depicted in Figure \ref{fig:pcolor}, the normalized table values of dimension meter were multiplied by \qty{850}{\gram\centi\meter\per\meter} to arrive at the desired dimension of gram times centimeter. The value 8.5 refers to the mass of injected \cotwo in gram.

\sisetup{table-number-alignment=center, round-mode=places, round-precision=2, scientific-notation=true, output-exponent-marker=\text{e}, retain-zero-exponent=true}

\subsection{24 hours}
\printdistances{distances_24h}

\subsection{48 hours}
\printdistances{distances_48h}

\subsection{72 hours}
\printdistances{distances_72h}

\subsection{96 hours}
\printdistances{distances_96h}

\subsection{120 hours}
\printdistances{distances_120h}

\section{Sparse data provided by the participants}\label{sec:sparse_data}
The following tables present the sparse data as provided by the participants. The values \ensuremath{P_{10}(\bar x)}, \ensuremath{P_{50}(\bar x)} and \ensuremath{P_{90}(\bar x)} indicate the P10, P50 and P90 values of the expected mean of the respective quantity, whereas \ensuremath{P_{10}(\sigma)}, \ensuremath{P_{50}(\sigma)} and \ensuremath{P_{90}(\sigma)} refer to the correspondingly expected standard deviation. The values are also contained in the respective participant repositories.

\sisetup{table-number-alignment=center, round-mode=places, round-precision=2, scientific-notation=true, output-exponent-marker=\text{e}, retain-zero-exponent=true}

\subsection{Austin}
\printsparsedata{austin}

\subsection{CSIRO}
\printsparsedata{csiro}

\subsection{Delft-DARSim}
\printsparsedata{delft-darsim}

\subsection{Delft-DARTS}
\printsparsedata{delft-darts}

\subsection{Heriot-Watt}
\printsparsedata{heriot-watt}

\subsection{LANL}
\printsparsedata{lanl}

\subsection{Melbourne}
\printsparsedata{melbourne}

\subsection{Stanford}
\printsparsedata{stanford}

\subsection{Stuttgart}
\printsparsedata{stuttgart}

\end{appendices}

\end{document}